# Nonlinear Optical Response Properties of Aryl - Substituted Boron - Dipyrromethene Dyes: Unidirectional Charge Transfer coupled with Structural Tuning

Ramprasad Misra[1]*
*Department of Physical Chemistry*
*Indian Association for the Cultivation of Science,*
*Jadavpur, Kolkata 700032, India*

## Abstract

Controlling the nonlinear optical (NLO) response properties at the molecular level is a key to develop strong NLO active materials for technological applications. In this paper, we report quantum chemical investigation of NLO response properties of select aryl-substituted Boron-Dipyrromethene (BODIPY) dyes, a class of intramolecular charge transfer (ICT) probes. Density functional theory (DFT) with long-range corrected CAM-B3LYP functional and cc-pVTZ, 6-31G(d,p) and 6-31+G(d,p) basis sets are employed to compute the electronic structures and NLO response of the aforesaid molecules. Calculations at the second order Møller-Plesset perturbation (MP2) level of theory are performed for comparison. The results suggest that the charge transfer process in these molecules is mostly unidirectional and the total first hyperpolarizability ($\beta_{total}$) values of these molecules are dominantly dictated by the response in the direction of charge transfer. Alteration of conjugation strength through donor/acceptor substitution as well as twisting of the phenyl ring obtained through incorporation of methyl groups affect the NLO response of thesemolecules. The vector components of first hyperpolarizability ($\beta_{vec}$) of the probe molecules are also studied to analyze the angle between the vector components of $\beta_{vec}$ and the dipole moment vector. The results presented here are expected to shed light on the origin of NLO response of several aryl-substituted BODIPY dyes and provide means to optimizing it for future technological applications.

[1]Present address: Department of Organic Chemistry, Weizmann Institute of Science, Rehovot 76100, Israel; *Corresponding author; E-mail: ramprasadmisra@gmail.com

**Introduction**

Studies of materials with strong nonlinear optical (NLO) response have been primarily driven by their potential for technological applications, including, in the photochromic switches, in three dimensional optical data storage, in fluorescence imaging and in telecommunications, among several others[1-7]. In recent years, intramolecular charge transfer (ICT) based organic molecules become most popular for these applications, owing to high flexibility, durability as well as ease to syntheses[8-13]. The intramolecular charge transfer (ICT) reaction of several substituted Boron-Dipyrromethene (BODIPY) dyes has been studied widely[14-15], but little seems to be known about their nonlinear optical (NLO) response properties. BODIPY dyes are strongly absorbing and generally show relatively sharp fluorescence emission with high quantum yields[16]. The absorption and emission properties of these molecules can be tuned conveniently by changing the substitution pattern of the BODIPY framework that can ultimately push their fluorescence into the near-infrared (NIR) region[16-18]. These dyes are reported to have excellent thermal and photochemical stability and the triplet-state formation is also negligible[16]. Very recently, some experimental studies on the NLO response of fluoroalkylated and dimethylaminostyryl substituted BODIPY dyes have been reported literature[19-20]. Lager et al[21] have reported the synthetic methodologies of several aryl-substituted BODIPY dyes. Incorporation of donor/ acceptor group in the phenyl ring in conjugation with the BODIPY can be used to affect the ICT process, ultimately altering the NLO response[22-24]. The donor/ acceptor substitution also helps to tune the energy gap between the highest occupied molecular orbital (HOMO) and lowest unoccupied molecular orbital (LUMO)[22,25-26]. Incorporation of the donor/ acceptor group is sometimes used to achieve non-centrosymmetry in several molecules[27], a criterion for obtaining non-vanishing first hyperpolarizability. As the NLO response properties of a material is ultimately shaped by the properties of the individual chromophores, tuning of the NLO response at the molecular level using the aforesaid strategies could be a key develop novel NLO materials.

Electron correlation plays a crucial role in the computation of NLO response properties of a molecule[28-29]. To keep a balance between the computational cost and accuracy, a Coulomb-attenuated hybrid exchange-correlation DFT functional, CAM-B3LYP is used for optimization of the structures. This functional is reported to be useful for computing properties of several charge transfer based molecules, including, their NLO response[30-32]. Calculations at the MP2 level of theory are performed to compute the electronic structures and NLO response of the selected probe molecules for comparison. The total first hyperpolarizability or intrinsic hyperpolarizability ($\beta_{total}$) is used widely to estimate the charge transfer from the donor to acceptor[9,13]. The values of $\beta_{total}$ are always positive, irrespective of the sign of the individual tensorial components and it cannot be measured experimentally. On the other hand, the vector component of the first hyperpolarizability ($\beta_{vec}$) can be measured by electric field induced second harmonic generation (EFISH) or Hyper-Rayleigh scattering (HRS) techniques[13]. As $\beta_{vec}$ could be positive or negative, it is expected to carry more information about the hyperpolarization process in the molecule than $\beta_{total}$. The ratio of $\beta_{vec}$ and $\beta_{total}$ also provides important information about the direction of charge transfer in the molecules as[13]

$$\frac{\beta_{vec}}{\beta_{total}} = cos\theta \quad (i)$$

where, θ is the angle between the vector formed by $\beta_{vec}$ components and the dipole moment vector.

In this paper, our aim is to understand the origin of NLO response of some novel aryl-substituted BODIPY dyes. The effect of change in structure through incorporation of methyl groups that restricts the motion of the phenyl ring and modulation of π-conjugation strength through wax and wane in the donor/ acceptor strength on the NLO response properties of these molecules are investigated. The methodologies adopted for this work are described in Section 2. Sections 3(a-b) deal with the ground state electronic structures and the absorption spectra of the molecules studied. The effect of change in electronic structures on the polarizability and first hyperpolarizability of the BODIPY dyes are discussed in Section 3c. A detailed analysis of $\beta_{vec}$ and $\beta_{total}$ values of the molecules studied is done to understand the directions of charge transfer in these molecules. How change in HOMO – LUMO energy gap affect the NLO response of the aforesaid molecules are also discussed in this section.

## 2. Methods

The ground state equilibrium geometries of the molecules under investigation (Scheme - 1) have been optimized fully using density functional theory (DFT) using CAM-B3LYP functional employing 6-31G(d,p), 6-31+G(d,p) and cc-pVTZ basis sets. The geometry optimizations are followed by calculations of components of polarizability (α) and the first hyperpolarizability (β) at the respective optimized structures at the same level of theory. The values of β have been reported to be quite sensitive on the basis set used. We have used Pople's split-valence 6-31G(d,p), 6-31+G(d,p) basis sets and Dunning's correlation-consistent cc-pVTZ basis set to study the effect of basis sets in these calculations. The effect of electron-correlation, which is reported to be an important parameter in polarizability calculations, has been taken care of using second order Møller - Plesset perturbation theory (MP2) calculations[28-29]. In every case, the lack of imaginary frequencies in the optimized structures confirmed the optimization of the geometries of the molecules. Time-dependent density functional theory (TDDFT) calculations were performed using CAM-B3LYP functional and cc-pVTZ basis set to get first ten vertical excited states of the molecules. All these calculations have been carried out with Gaussian09 quantum chemistry software[33], while the visualizations of the structures have been carried out by Chemcraft[34] and GausView 5.0[35] softwares.

The dipole moment ($\mu$) of the molecules under investigation has been calculated using the following equation[12] –

$$\mu = (\mu_x^2 + \mu_y^2 + \mu_z^2)^{1/2} \qquad \text{(ii)}$$

where, $\mu_x$, $\mu_y$ and $\mu_z$ are components of the dipole moments in the $x$, $y$ and $z$ directions, respectively. The average polarizabilities ($\alpha_{av}$) of the molecules are obtained from the components of polarizability using the following equation[12, 36]:

$$\alpha_{av} = \frac{1}{3}(\alpha_{xx} + \alpha_{yy} + \alpha_{zz}) \qquad \text{(iii)}$$

The first hyperpolarizability ($\beta$) is a tensor of rank 3 with twenty seven components that are reduced to ten by the virtue of Kleinman symmetry of the ($3 \times 3 \times 3$) matrix representing the $\beta$ tensor[31]. The values of total first hyperpolarizability ($\beta_{total}$) of the molecules (1-6) are calculated using the $x$, $y$ and $z$ components of first hyperpolarizability using the equation[12,36] (iv).

$$\beta_{total} = \left[ \beta_x^2 + \beta_y^2 + \beta_z^2 \right]^{1/2} \qquad \text{(iv)}$$

where, $\beta_i = \beta_{iii} + \sum_{i \neq j} \left( \beta_{ijj} + \beta_{jij} + \beta_{jji} \right) \quad (i = x, y, z)$ \qquad (v)

Using Kleinman symmetry[12], we get $\beta_x = \left( \beta_{xxx} + \beta_{xyy} + \beta_{xzz} \right)$ \qquad (vi)

Similarly, $\beta_y = \left( \beta_{yyy} + \beta_{yxx} + \beta_{yzz} \right)$ and $\beta_z = \left( \beta_{zzz} + \beta_{zyy} + \beta_{zxx} \right)$.

The values of $\beta_{vec}$ at the static limit can be calculated from the tensorial components of first hyperpolarizability using equation (vii)[12,13].

$$\beta_{vec} = \frac{\left( \beta_x \mu_x + \beta_y \mu_y + \beta_z \mu_z \right)}{\mu} \qquad \text{(vii)}$$

## 3. Results and Discussion

### 3a. The electronic structure of boron dipyrromethene dyes (1 - 6)

The electronic structure of a molecule plays a crucial role in determining its properties. The structure, therefore, the NLO response property of a molecule is very sensitive to the theory and basis set used. We have used DFT (CAM-B3LYP functional) and MP2 level of theories to optimize the ground state structures of the BODIPY dyes, used for our studies. The front and side views optimized structures of molecules (1-2) using DFT (CAM-B3LYP functional) and cc-pVTZ basis set are displayed in figure 1. The structures of molecules (3-6) obtained using same level of theory are shown in figure 2. The important geometrical parameters have been reported in Table 1. The ground state optimized structures obtained using DFT (CAM-B3LYP functional) and MP2 level of theories employing 6-31G(d,p) and 6-31+G(d,p) basis sets are shown in figures (S1-S4) in the Supporting Information. The results show that the 3-4-5-6 dihedral angle (Scheme - 1) gets twisted due to incorporation of the methyl groups. For example, CAM-B3LYP/cc-pVTZ calculations predict the aforesaid dihedral angle to be 90° for molecules 1, 5 and 6 vis-à-vis 57°, 50° and 60°, for molecules 2, 3 and 4, respectively. The predicted values of 3-4-5-6 dihedral angle are comparable in all other theory and basis set combinations. The changes in predicted bond lengths of the molecules due to substitution of methyl groups and also due to donor/ acceptor substitution are rather small (Table - 1). The changes in structures of aryl-substituted BODIPY dyes affect several properties of the molecules. For example, Tang and co-workers[14] have reported that attaching an unsubstituted phenyl ring to the BODIDY (molecule 2, scheme - 1) decreases the quantum yield of fluorescence ($\phi_f$) compared to BODIPY alone, incorporation of two methyl groups (molecule 1, Scheme - 1) in the BODIPY moiety restricts the motion of the phenyl ring, ultimately increasing the $\phi_f$ of the molecule. Effect of the change in the ground state electronic structures on the absorption spectra, polarizabilities and first hyperpolarizabilities of the aryl-substituted BODIPY dyes are discussed in the following sections.

### 3b. Molecular orbital picture and absorption spectra of the BODIPY dyes

Time dependent density functional theory (TDDFT) calculations were performed on the ground state optimized geometries of the molecules (1-6) obtained using CAM-B3LYP functional and cc-pVTZ basis set to get the energy and oscillator strength of the first ten vertical excitation of these molecules. The vertical transition maxima, oscillator strength first two excited states and contribution from the dominant transitions have been reported in Table 2. The results show that $S_0 \rightarrow S_1$ transitions dominantly dictate the electronic absorption profiles of the molecules studied, except for molecule 3 where both $S_0 \rightarrow S_1$ and $S_0 \rightarrow S_2$ transitions are expected to occur. The computed energy ($\Delta E_{gs}$) and oscillator strength ($f$) of $S_0 \rightarrow S_1$ transition in molecule 1 are 410.35 nm and 0.535, respectively. The $\Delta E_{gs}$ and $f$ values for molecule 2 are 397.69 nm and 0.441, respectively. The $\Delta E_{gs}$ value gets slightly blue shifted due to substitution of –N(CH$_3$)$_2$ group, while substitution of –NO$_2$ group leads to slight red shift from their unsubstituted counterparts. The oscillator strength values of higher excited states of these molecules are too low to contribute to their absorption spectra. The highest occupied molecular orbital (HOMO) and lowest unoccupied molecular orbital (LUMO) pictures of molecules (1-6) obtained using CAM-B3LYP functional and cc-pVTZ basis set are shown in figure 4. TDDFT results (Table - 2) indicate that HOMO to LUMO transitions dominantly dictate the absorption spectra of the molecules studied, except for molecule 3, where (HOMO-1) to LUMO transition also plays an important role in shaping its absorption profile.

### 3c. Linear and nonlinear optical response properties of molecules

The components of polarizability and first hyperpolarizability of the molecules under investigation are computed using their respective optimized geometries. The average polarizability ($\alpha_{av}$) and total first hyperpolarizability ($\beta_{total}$) are obtained from their components using equation (iii) and (iv), respectively are reported in Table - 3. The dipole moment ($\mu$) values of these molecules are also reported in the same table. The results show that the dipole moment in the $x$-direction ($\mu_x$) dominantly dictate the $\mu$ values of the molecules, except for molecules 4 and 6 (Table – S1 in Supporting Information). The dipole moment is highest in molecule 3 while that is lowest in molecule 4. The change (in percentage) in the values of $\alpha_{av}$ are rather less than that of $\mu$. The values of $\mu$, $\alpha_{av}$ and $\beta_{total}$ of obtained using DFT and MP2 level of theory with different basis sets are shown in figure 5(a-c) for comparison. The figures show that the trend in values of $\mu$, $\alpha_{av}$ as well as $\beta_{total}$ of molecules (1-6) seems to be unchanged due to the theory/ basis sets used for our calculations. The values of $\mu$, $\alpha_{av}$ and $\beta_{total}$ of molecules studied are seen to increase due to use of 6-31+G(d,p) basis set, compared to that obtained using 6-31G(d,p) basis set. The changes in structure affect the NLO response of the BODIPY dyes. An increase in values in $\beta_{total}$ is observed in molecule 1 (1130.48 a.u. vis-à-vis 483.07 a.u. in molecule 2) due to incorporation of two methyl groups in the BODIPY moiety. The change in conjugation strength due to incorporation of donor/ acceptor group affects the $\beta_{total}$ values. The computed value of $\beta_{total}$ of molecule 2 is 483.07 a.u. that is increased to 3942.95 a.u. and 1135.57 a.u. due to substitution of the $-N(CH_3)_2$ (molecule 3) and $-NO_2$ (molecule 4) groups, respectively. Incorporation of $-NO_2$ (molecule 6) group in molecule 1 shows an increase in the $\beta_{total}$ values while addition of $-N(CH_3)_2$ group in the same molecule seems to be detrimental. The HOMO – LUMO band gap ($\Delta E_{H-L}$) of these molecules are calculated and reported in the Table – S2 in Supporting Information. Generally, the values of first hyperpolarizability increases with decrease in $\Delta E_{H-L}$[26]. The values of $\beta_{total}$ increase with the decrease of $\Delta E_{H-L}$, except for molecule 3. This deviation is expected as HOMO to LUMO as well as (HOMO-1) to LUMO transitions shape the absorption profile of molecule 3 (Table - 2), unlike other molecules studied, where HOMO to LUMO transition dominantly shape their absorption behavior. Yu and co-workers[37] have reported that $\beta_y$ dominantly dictate the values of the $\beta_{total}$ of some metal cationic complex of 2,3-naphtho-15-crown-5 ether derivatives, leading them to conclude that during polarization process the charge is expected to transfer in the $y$-direction. The $\beta_x$, $\beta_y$ and $\beta_z$ values are calculated from the tensorial components of hyperpolarizability are reported in Table – S3 in Supporting Information. The results show that the $\beta_x$ dominantly dictate the values of $\beta_{total}$ for all the molecules studied, while the contribution of $\beta_y$ and $\beta_z$ to $\beta_{total}$ are quite small. This suggests that during polarization process of molecules (1-6) the charge is expected to transfer in the *x*-direction. In strong donor – acceptor system, where charge transfer is favored, the ICT is unidirectional and parallel to the molecular dipole moment, so that one of the components accounts for almost all of the NLO response[13,38]. So, one can expect maximum charge delocalization when the ratio of $\beta_{vec}$ and $\beta_{total}$ is unity[38]. The values of $\beta_{vec}$ as obtained using equation (vii) from the tensorial components of hyperpolarizability are reported in Table – 4. The absolute values of $\beta_{vec}$ of obtained using DFT and MP2 level of theory with different basis sets are shown in figure 5d for comparison. The ratio of $\beta_{vec}$ to $\beta_{total}$ is almost unity in most of the cases, except for molecules 4 and 6 that has $-NO_2$ substitution. This ratio of $\beta_{vec}$ to $\beta_{total}$ implies the unidirectional and parallel charge transfer from the BODIPY moiety to aryl ring. Wax and wane of either $\beta_{total}$ or $\beta_{vec}$ values due to substitution of $-N(CH_3)_2/-NO_2$ group also implies that the direction of charge transfer also influence the NLO response of these molecules. To the best of our knowledge, this is the first report on unidirectional charge transfer in aryl-substituted BODIPY dyes. As controlling the change transfer in the molecular scale is a key to design molecular devices for nanooptoelectronics[39], these results are expected to draw attention to scientific community for further studies of molecular electronics.

**Conclusion**

Quantum chemical calculations at DFT (CAM-B3LYP functional) and MP2 level of theories are performed to study the electronic structure and NLO response of some novel aryl-substituted BODIPY dyes. The trends of values of dipole moment, average polarizability and first hyperpolarizability of the molecules studied are consistent at the level of theory and basis set combinations used for our calculations. The dipole moment ($\mu$) and first hyporpolarizability ($\beta_{total}$) of these molecules are dominantly dictated by their response in the $x$-direction, indicating strong unidirectional charge transfer in the molecules studied. The ratio of vector components of first hyperpolarizability ($\beta_{vec}$) to that of $\beta_{total}$ is close to unity, except for molecules with –NO$_2$ substitution that also supports this behavior. The unidirectional charge transfer and the structural changes affect the NLO response of the molecules studied. As controlling the change transfer in the molecular scale is a key to design molecular devices for technological applications, the insight obtained from these studies are expected to be useful in designing novel BODIPY dyes for nonlinear optical applications.

**Acknowledgements**

The author is grateful to Prof. S.P. Bhattacharyya for encouraging him in taking up this work and Prof. D.S. Ray, IACS for several fruitful discussions. The Computer Centre of IACS acknowledged for providing with the computational facilities. The author would like to thank Dr. Pralok K. Samanta and Dr. Debasree Manna (WIS, Israel) for technical helps.

**Supporting Information**

The $x$, $y$ and $z$ components of dipole moments (Debye) and first hyperpolarizability (a.u.);

The ground state optimized structures obtained using DFT (CAM-B3LYP functional) and

MP2 level of theories employing 6-31G(d,p) and 6-31+G(d,p) basis sets.

**Figure Captions**

Figure 1: The front and side views of ground state optimized geometries of molecule 1 (panel a - b) and molecule 2 (panel c - d), obtained using DFT (CAM-B3LYP functional) calculations with cc-pVTZ basis set. The $x$, $y$ and $z$ directions are shown for reference. The dihedral angle is $90^0$ in molecule 1 vis-à-vis $57^o$ in molecule 2 due to the presence of methyl groups in the BODIPY moiety of the former.

Figure 2: The ground state optimized geometries of molecules (3 - 6) as obtained using DFT, employing CAM-B3LYP functional and cc-pVTZ basis set. The dihedral angle is changed due to incorporation of methyl groups in molecules 5 and 6, compared to molecules 3 and 4, respectively. The $x$, $y$ and $z$ directions are shown for reference.

Figure 3: The highest occupied molecular orbital (HOMO) and lowest unoccupied molecular orbital LUMO pictures of molecules (a) 1, 5 and 6 and (b) molecules 2, 3 and 4, obtained using DFT, employing CAM-B3LYP functional and cc-pVTZ basis set..

Figure 4:  Comparison of (a) dipole moment ($\mu$), (b) average polarizability ($\alpha_{av}$), (c) first hyperpolarizability ($\beta_{total}$) and (d) absolute values of vector component of first hyperpolarizability ($/\beta_{vec}/$) of molecules (1 - 6) as obtained using DFT (CAM-B3LYP functional) employing cc-pVTZ, 6-31G(d,p) and 6-31+G(d,p) basis sets. The $\mu$, $\alpha_{av}$, $\beta_{total}$ and $/\beta_{vec}/$ values of molecules (2 - 4) using MP2 employing 6-31G(d,p) and 6-31+G(d,p) basis sets ate also presented for comparison.

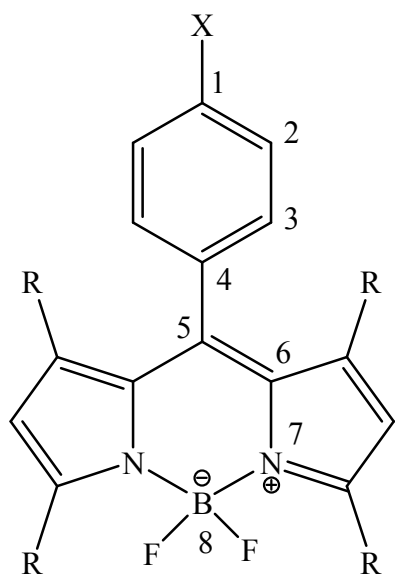

**Scheme - 1: The structure of the BODIPY dyes under investigation. For Molecule 1, R= -CH₃, X= -H; for molecule 2, R= -H, X= -H; for molecule 3, for molecule 2, R= -H, X= - N(CH₃)₂; for molecule 4, R= -H, X= -NO₂; for molecule 5, R= - CH₃, X= - N(CH₃)₂ and for molecule 6, R= - CH₃, X= - NO₂.**
**Numbering of some of the heavy atoms are done for reference.**

**Table 1a: The bond lengths (Å) and dihedral angle (in degree) of the molecule (1-6) as obtained using DFT (CAM-B3LYP functional) employing cc-pVTZ, 6-31G(d,p) and 6-31+G(d,p) basis sets.**

| Theory/ Molecule | 1-2 | 2-3 | 3-4 | 4-5 | 5-6 | 6-7 | 7-8 | 3-4-5-6 |
|---|---|---|---|---|---|---|---|---|
| DFT/ cc-pVTZ | | | | | | | | |
| 1 | 1.385 | 1.385 | 1.388 | 1.487 | 1.392 | 1.393 | 1.549 | 90 |
| 2 | 1.384 | 1.383 | 1.392 | 1.477 | 1.394 | 1.385 | 1.559 | 57 |
| 3 | 1.405 | 1.376 | 1.394 | 1.467 | 1.399 | 1.384 | 1.557 | 50 |
| 4 | 1.380 | 1.381 | 1.392 | 1.479 | 1.391 | 1.385 | 1.561 | 50 |
| 5 | 1.403 | 1.380 | 1.386 | 1.485 | 1.393 | 1.393 | 1.548 | 90 |
| 6 | 1.380 | 1.382 | 1.389 | 1.486 | 1.391 | 1.393 | 1.550 | 90 |
| DFT/6-31G(d,p) | | | | | | | | |
| 1 | 1.390 | 1.390 | 1.394 | 1.491 | 1.397 | 1.395 | 1.550 | 90 |
| 2 | 1.390 | 1.388 | 1.398 | 1.480 | 1.400 | 1.388 | 1.560 | 54 |
| 3 | 1.410 | 1.382 | 1.400 | 1.470 | 1.405 | 1.388 | 1.560 | 48 |
| 4 | 1.386 | 1.386 | 1.399 | 1.482 | 1.398 | 1.388 | 1.562 | 56 |
| 5 | 1.408 | 1.386 | 1.393 | 1.489 | 1.398 | 1.395 | 1.549 | 90 |
| 6 | 1.386 | 1.387 | 1.396 | 1.491 | 1.397 | 1.395 | 1.551 | 90 |
| DFT/6-31+G(d,p) | | | | | | | | |
| 1 | 1.392 | 1.391 | 1.395 | 1.491 | 1.398 | 1.398 | 1.546 | 90 |
| 2 | 1.392 | 1.390 | 1.399 | 1.481 | 1.400 | 1.390 | 1.555 | 57 |
| 3 | 1.412 | 1.384 | 1.400 | 1.471 | 1.405 | 1.389 | 1.553 | 51 |
| 4 | 1.387 | 1.388 | 1.399 | 1.484 | 1.398 | 1.390 | 1.558 | 59 |
| 5 | 1.409 | 1.388 | 1.393 | 1.489 | 1.398 | 1.398 | 1.546 | 90 |
| 6 | 1.388 | 1.388 | 1.396 | 1.491 | 1.398 | 1.398 | 1.548 | 90 |

**Table 1b: The bond lengths (Å) and dihedral angle (in degree) of the molecule (2-4) as obtained using MP2 theory and 6-31G(d,p) and 6-31+G(d,p) basis sets.**

| Theory/ Molecule | 1-2 | 2-3 | 3-4 | 4-5 | 5-6 | 6-7 | 7-8 | 3-4-5-6 |
|---|---|---|---|---|---|---|---|---|
| MP2/6-31G(d,p) | | | | | | | | |
| 2 | 1.397 | 1.394 | 1.405 | 1.471 | 1.403 | 1.384 | 1.566 | 52 |
| 3 | 1.413 | 1.389 | 1.405 | 1.463 | 1.406 | 1.384 | 1.564 | 49 |
| 4 | 1.393 | 1.392 | 1.406 | 1.472 | 1.403 | 1.384 | 1.567 | 53 |
| MP2/6-31+G(d,p) | | | | | | | | |
| 2 | 1.399 | 1.397 | 1.406 | 1.473 | 1.403 | 1.386 | 1.562 | 57 |
| 3 | 1.415 | 1.392 | 1.406 | 1.464 | 1.406 | 1.386 | 1.559 | 52 |
| 4 | 1.395 | 1.395 | 1.406 | 1.473 | 1.403 | 1.386 | 1.563 | 58 |

**Table 2: The vertical transition maxima (in nm) and their corresponding oscillator strengths of molecules (1-6), as predicted from TDDFT calculations employing CAM-B3LYP functional and cc-pVTZ basis set. H and L stand for HOMO and LUMO respectively.**

| Molecule/ Excited state | Transition maxima (nm) | Oscillator strength | Dominant transitions[a] |
|---|---|---|---|
| 1 $S_1$ | 410.35 | 0.535 | H – L (0.70) |
| $S_2$ | 315.48 | 0.047 | (H-2) – L (0.70) |
| 2 $S_1$ | 397.69 | 0.441 | H – L (0.69) |
| $S_2$ | 309.71 | 0.075 | (H-1) – L (0.69) |
| 3 $S_1$ | 394.47 | 0.4238 | H – L (0.69) |
| $S_2$ | 382.42 | 0.3393 | (H-1) – L (0.69) |
| 4 $S_1$ | 403.69 | 0.430 | H – L (0.68) |
| $S_2$ | 314.16 | 0.067 | (H-1) – L (0.67) |
| 5 $S_1$ | 409.83 | 0.526 | H – L (0.70) |
| $S_2$ | 351.29 | 0.00 | (H-1) – L (0.69) |
| 6 $S_1$ | 412.29 | 0.536 | H – L (0.70) |
| $S_2$ | 333.15 | 0.00 | H – (L+1) (0.69) |

[a]The contribution from corresponding transitions are mentioned in the parenthesis.

**Table 3a: The values of dipole moments (μ, Debye), $\alpha_{av}$ (a.u.) and $\beta_{total}$ (a.u.) of Molecules (1-6) as calculated by using density functional theory (DFT) with CAM-B3LYP functional employing cc-pVTZ, 6-31G(d,p) and 6-31+G(d,p) basis sets.**

| Molecule | cc-pVTZ | | | 6-31G(d,p) | | | 6-31+G(d,p) | | |
|---|---|---|---|---|---|---|---|---|---|
| | μ | $\alpha_{av}$ | $\beta_{total}$ | μ | $\alpha_{av}$ | $\beta_{total}$ | μ | $\alpha_{av}$ | $\beta_{total}$ |
| 1 | 4.81 | 267.31 | 1130.48 | 4.72 | 245.17 | 1156.26 | 5.12 | 272.06 | 1415.31 |
| 2 | 5.64 | 212.91 | 483.07 | 5.48 | 195.04 | 427.29 | 6.01 | 220.73 | 534.86 |
| 3 | 9.25 | 264.74 | 3942.95 | 8.97 | 242.98 | 3993.22 | 9.68 | 271.59 | 4570.96 |
| 4 | 0.85 | 231.81 | 1135.57 | 0.83 | 213.7 | 1234.32 | 0.88 | 240.87 | 1359.28 |
| 5 | 7.43 | 309.29 | 645.08 | 7.28 | 283.35 | 576.49 | 7.75 | 312.71 | 690.28 |
| 6 | 1.10 | 286.58 | 1569.12 | 1.08 | 264.10 | 1714.79 | 1.18 | 292.78 | 1973.91 |

**Table 3b: The values of dipole moments (μ, Debye), $\alpha_{av}$ (a.u.) and $\beta_{total}$ (a.u.) of Molecules (2-4) as calculated by using MP2 theory, employing 6-31G(d,p) and 6-31+G(d,p) basis sets.**

| Molecule | 6-31G(d,p) | | | 6-31+G(d,p) | | |
|---|---|---|---|---|---|---|
| | μ | $\alpha_{av}$ | $\beta_{total}$ | μ | $\alpha_{av}$ | $\beta_{total}$ |
| 2 | 5.23 | 198.21 | 63.93 | 5.57 | 226.11 | 155.84 |
| 3 | 8.05 | 245.22 | 4001.33 | 8.44 | 275.55 | 4335.06 |
| 4 | 1.15 | 217.24 | 849.45 | 1.24 | 246.47 | 920.06 |

**Table 4a: The values of $\beta_x$ (a.u.) and $\beta_{vec}$ (a.u.) of Molecules (1-6) as calculated by using density functional theory (DFT) with CAM-B3LYP functional employing cc-pVTZ, 6-31G(d,p) and 6-31+G(d,p) basis sets.**

| Molecule | cc-pVTZ | | | 6-31G(d,p) | | | 6-31+G(d,p) | | |
|---|---|---|---|---|---|---|---|---|---|
| | $\beta_x$ | $\beta_{vec}$ | $\beta_{vec}$ /$\beta_{total}$ | $\beta_x$ | $\beta_{vec}$ | $\beta_{vec}$ /$\beta_{total}$ | $\beta_x$ | $\beta_{vec}$ | $\beta_{vec}$ /$\beta_{total}$ |
| 1 | 1130.48 | 1109.84 | 0.982 | 1156.26 | 1136.40 | 0.983 | 1415.31 | -1387.52 | -0.980 |
| 2 | 482.76 | 481.71 | 0.997 | 427.28 | 423.79 | 0.992 | 534.84 | 530.96 | 0.993 |
| 3 | -3942.9 | -3923.92 | -0.995 | -3992.63 | -3978.13 | -0.996 | -4570.73 | -4547.57 | -0.995 |
| 4 | 1135.56 | 518.33 | 0.456 | 1234.31 | 294.43 | 0.238 | 1359.28 | 496.28 | 0.365 |
| 5 | -641.9 | 644.12 | 0.998 | 576.25 | 572.38 | 0.993 | -680.61 | 689.04 | 0.998 |
| 6 | 1568.47 | -707.58 | -0.451 | -1714.75 | -911.67 | -0.53 | -1973.91 | -1019.83 | -0.517 |

**Table 4b: The values of $\beta_x$ (a.u.) and $\beta_{vec}$ (a.u.) of Molecules (2-4) as calculated by using MP2 theory, employing 6-31G(d,p) and 6-31+G(d,p) basis sets.**

| Molecule | 6-31G(d,p) | | | 6-31+G(d,p) | | |
|---|---|---|---|---|---|---|
| | $\beta_x$ | $\beta_{vec}$ | $\beta_{vec}$ /$\beta_{total}$ | $\beta_x$ | $\beta_{vec}$ | $\beta_{vec}$ /$\beta_{total}$ |
| 2 | 55.39 | 58.21 | 0.91 | 140.52 | 150.51 | 0.966 |
| 3 | -4000.11 | -3955.42 | -0.989 | -4331.45 | -4261.01 | -0.983 |
| 4 | 849.26 | 214.03 | 0.252 | 919.44 | 246.94 | 0.268 |

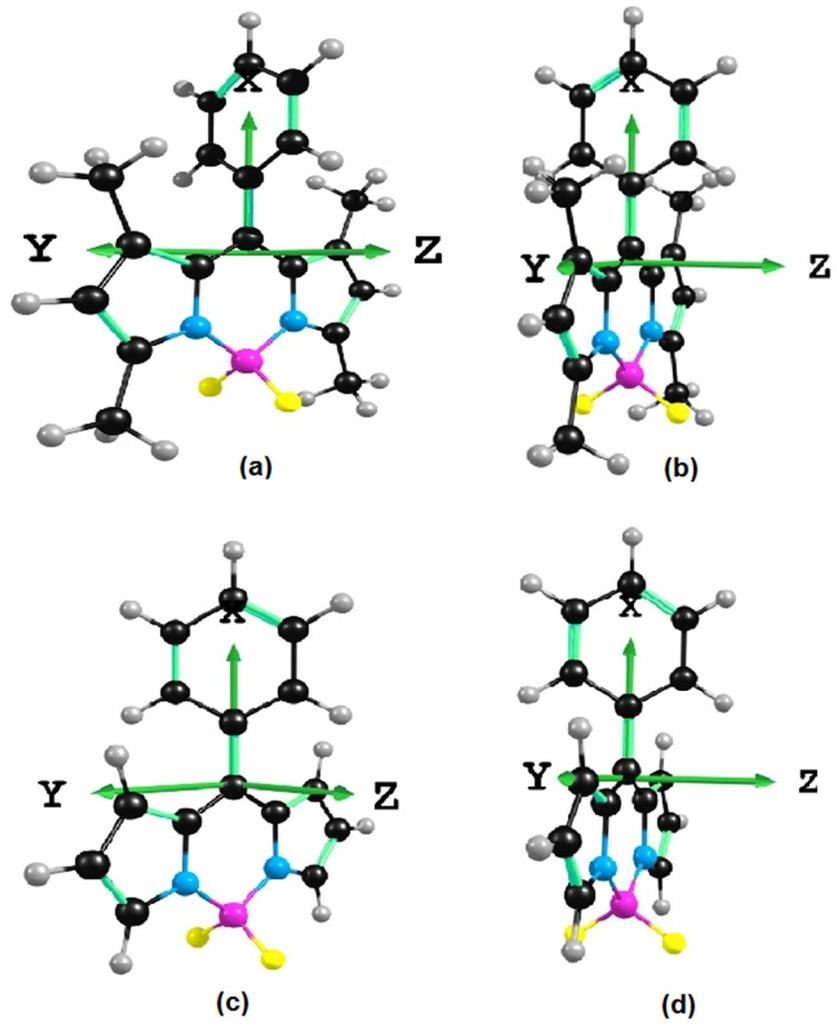

**Figure 1**

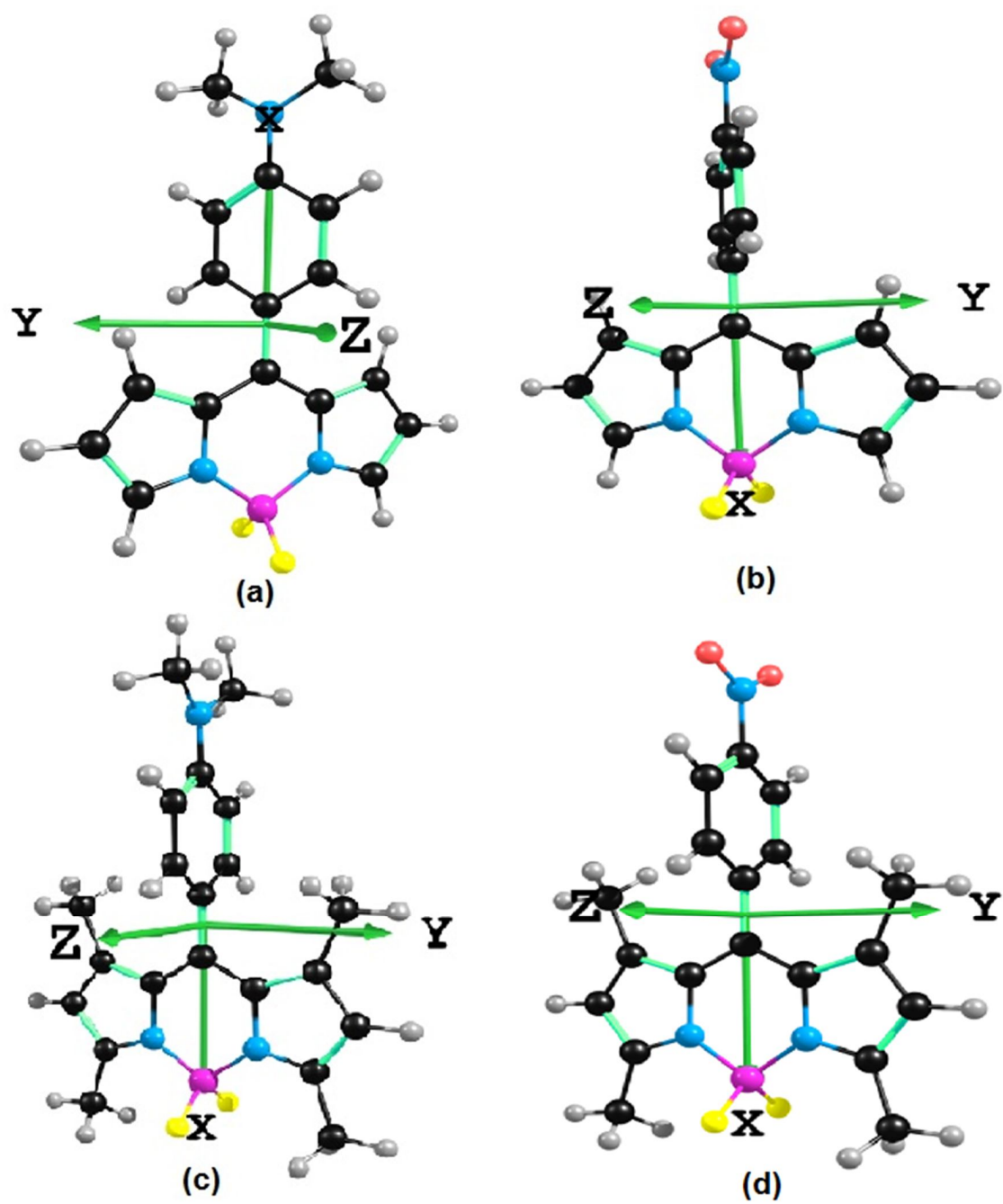

**Figure 2**

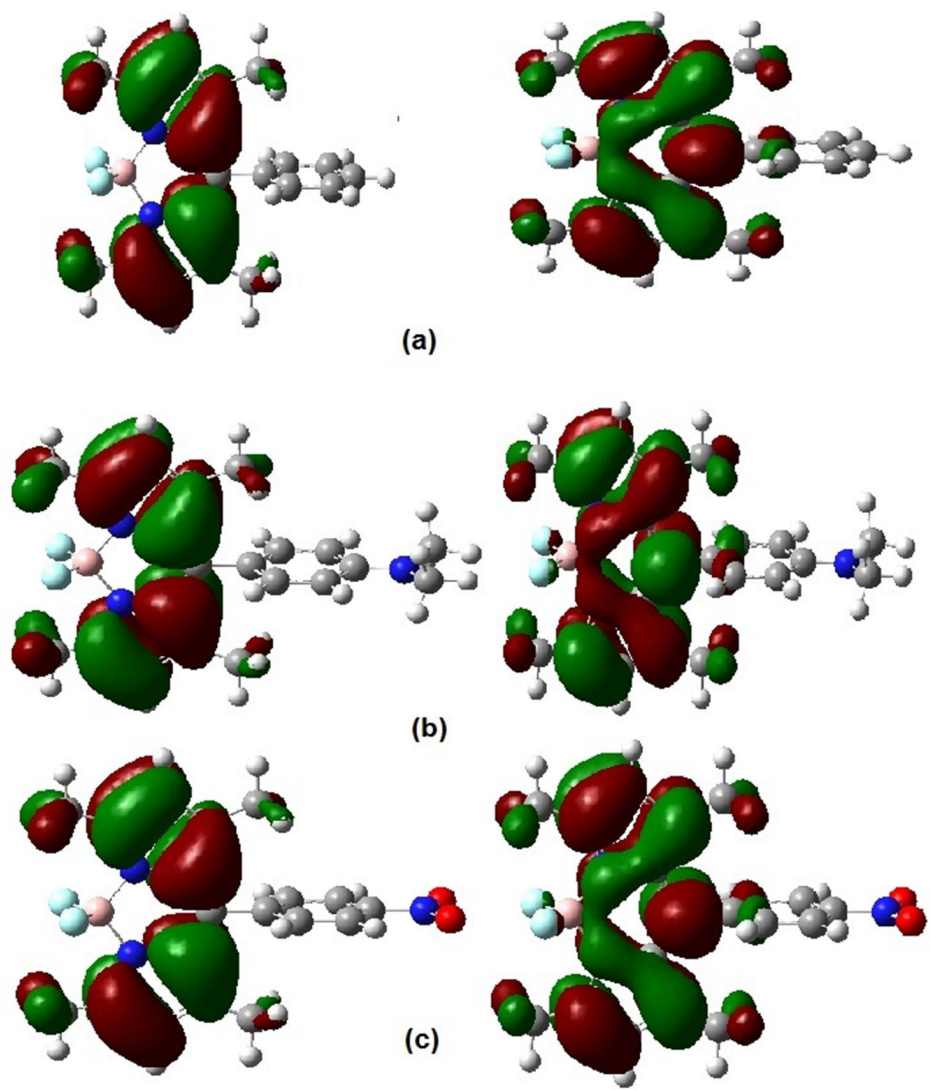

(a)

(b)

(c)

**Figure 3a**

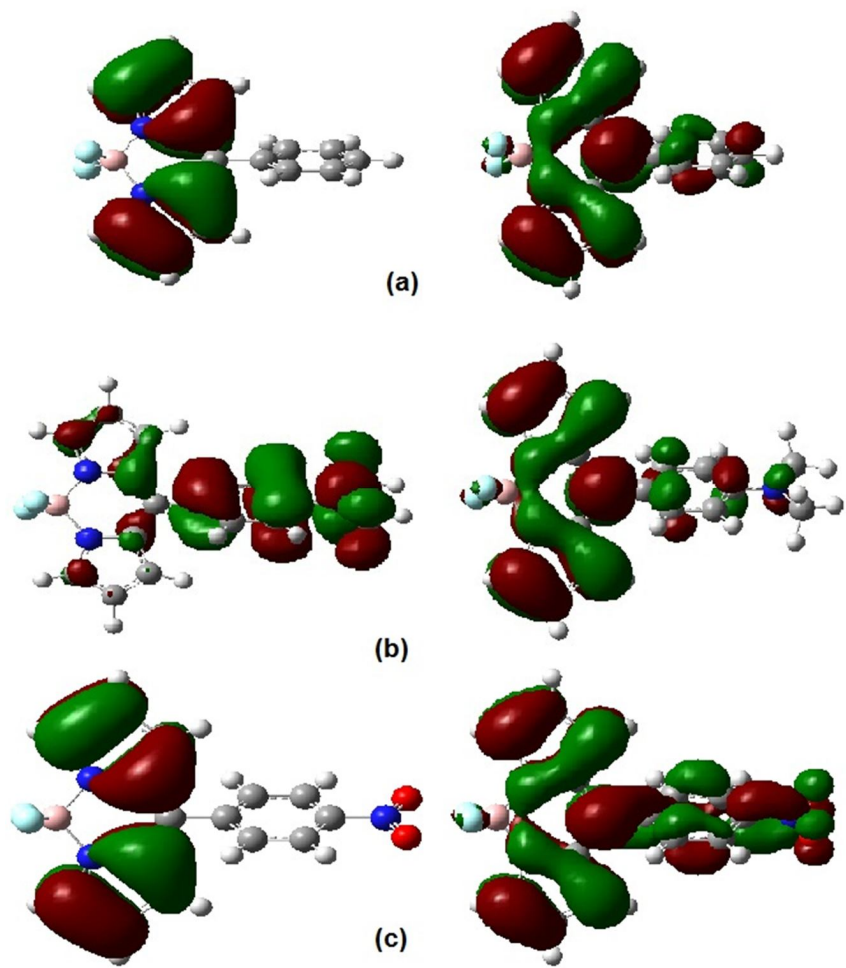

(a)

(b)

(c)

**Figure 3b**

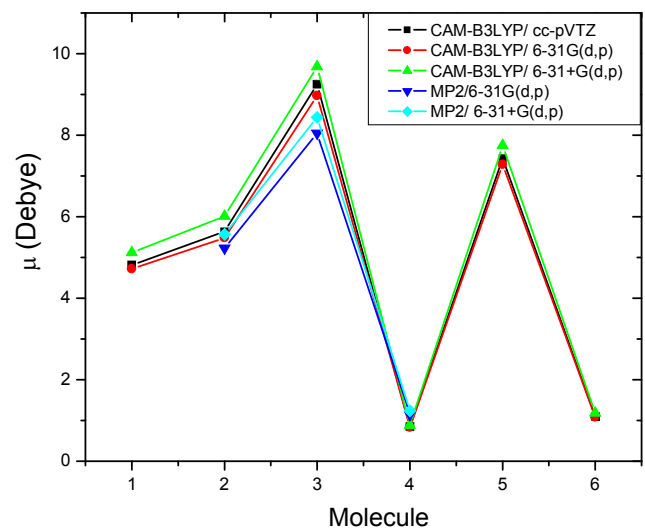

**Figure 4a**

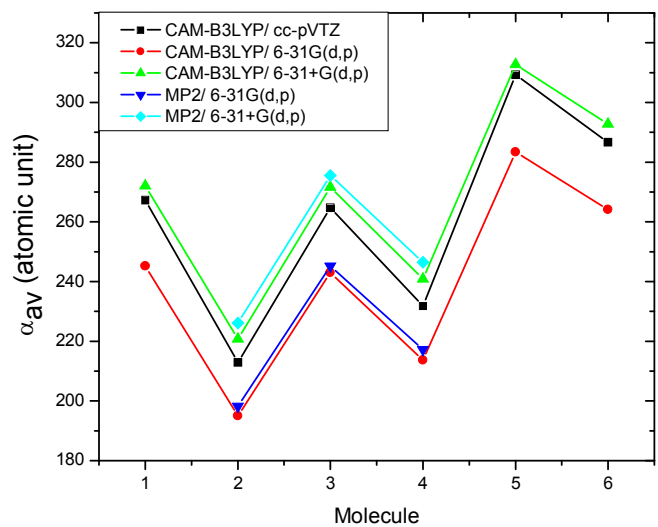

**Figure 4b**

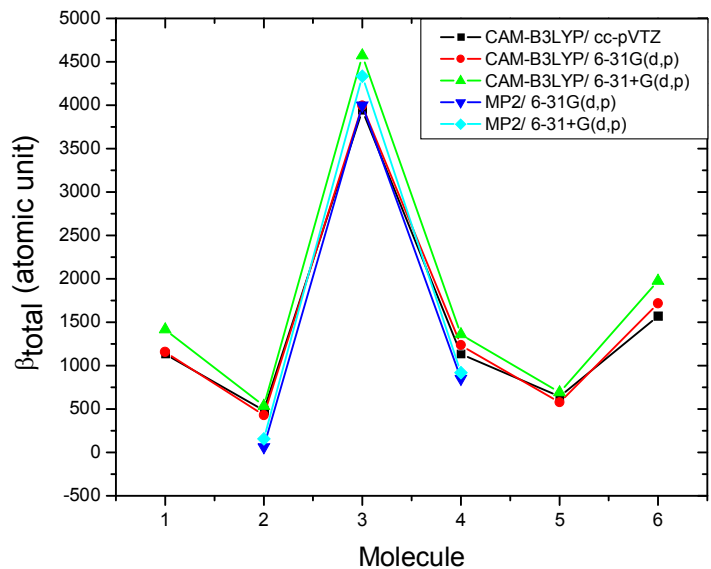

**Figure 4c**

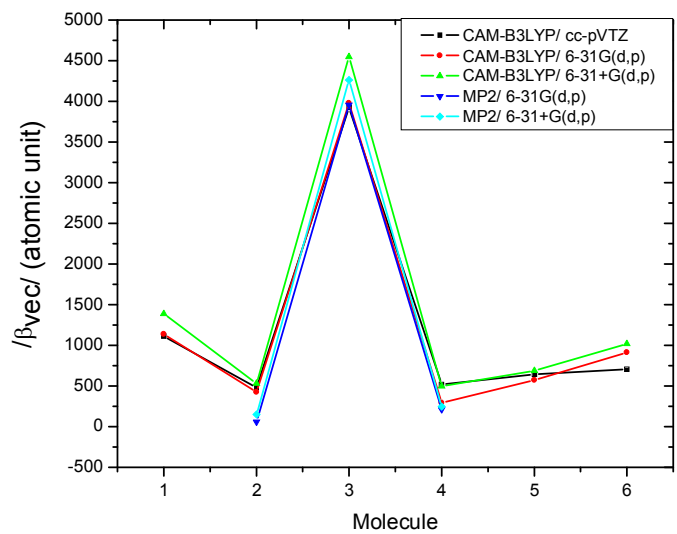

**Figure 4d**

*Supporting Information*

**Nonlinear Optical Response Properties of Aryl - Substituted Boron - Dipyrromethene Dyes: Unidirectional Charge Transfer coupled with Structural Tuning**

Ramprasad Misra[1]*

*Department of Physical Chemistry*

*Indian Association for the Cultivation of Science,*

*Jadavpur, Kolkata 700032, India*

**Figure Captions**

Figure S1: The ground state optimized geometries of molecules (1 - 6) as obtained using DFT, employing CAM-B3LYP functional and 6-31G(d,p) basis set. The $x$, $y$ and $z$ coordinates are shown for reference.

Figure S2: The ground state optimized geometries of molecules (1 - 6) as obtained using DFT, employing CAM-B3LYP functional and 6-31+G(d,p)basis set. The $x$, $y$ and $z$ coordinates are shown for reference.

Figure S3: The ground state optimized geometries of molecules (2 - 4) as obtained using MP2 theory and 6-31G(d,p) basis set. The $x$, $y$ and $z$ coordinates are shown for reference.

Figure S4: The ground state optimized geometries of molecules (2 - 4) as obtained employing MP2 theory and 6-31+G(d,p) basis set. The $x$, $y$ and $z$ coordinates are shown for reference.

---

[1]Present address: Department of Organic Chemistry, Weizmann Institute of Science, Rehovot 76100, Israel

**Table S1a: The x, y and z components of dipole moments (Debye) of molecules (1-6) as calculated using DFT (CAM-B3LYP functional), employing cc-pVTZ, 6-31G(d,p) and 6-31+G(d,p) basis sets.**

| Molecule | cc-pVTZ | | | 6-31G(d,p) | | | 6-31+G(d,p) | | |
|----------|---------|---------|---------|------------|---------|---------|-------------|---------|---------|
| | $\mu_x$ | $\mu_y$ | $\mu_z$ | $\mu_x$ | $\mu_y$ | $\mu_z$ | $\mu_x$ | $\mu_y$ | $\mu_z$ |
| 1 | 4.72 | 0.87 | 0.40 | 4.64 | 0.81 | 0.35 | -5.02 | 0.89 | 0.42 |
| 2 | 5.95 | 0.70 | 0.29 | 5.43 | 0.69 | -0.21 | 5.95 | 0.73 | 0.30 |
| 3 | 9.21 | 0.80 | 0.30 | 8.94 | 0.78 | -0.17 | 9.64 | 0.82 | 0.42 |
| 4 | 0.38 | 0.74 | 0.12 | 0.20 | 0.74 | -0.32 | 0.32 | 0.75 | 0.34 |
| 5 | -7.35 | 1.04 | 0.47 | 7.20 | 1.07 | 0.41 | -7.66 | 1.10 | 0.49 |
| 6 | -0.49 | 0.89 | 0.43 | 0.57 | 0.82 | 0.38 | 0.61 | 0.90 | 0.45 |

**Table S1b: Table S1a: The x, y and z components of dipole moments (Debye) of molecules (2-4) as calculated using MP2 theory, employing 6-31G(d,p) and 6-31+G(d,p) basis sets.**

| Molecule | 6-31G(d,p) | | | 6-31+G(d,p) | | |
|----------|------------|---------|---------|-------------|---------|---------|
| | $\mu_x$ | $\mu_y$ | $\mu_z$ | $\mu_x$ | $\mu_y$ | $\mu_z$ |
| 2 | 5.14 | 0.74 | -0.65 | 5.48 | 0.83 | -0.64 |
| 3 | 7.98 | 1.07 | -0.28 | 8.35 | 1.16 | -0.29 |
| 4 | 0.27 | 0.81 | -0.77 | 0.305 | 0.91 | -0.79 |

**Table S2: The HOMO – LUMO energy gap (eV) of molecules (1-6) as obtained using DFT calculation, employing CAM-B3LYP functional with cc-pVTZ and 6-31+G(d,p) basis sets. Corresponding $\beta_{total}$ values are also reported.**

| Molecule | cc-pVTZ | | 6-31+G(d,p) | |
|:---:|:---:|:---:|:---:|:---:|
| | Gap | $\beta_{total}$ | Gap | $\beta_{total}$ |
| 1 | 5.217 | 1130.48 | 5.166 | 1415.31 |
| 2 | 5.314 | 483.07 | 5.262 | 534.86 |
| 3 | 5.328 | 3942.95 | 5.256 | 4570.96 |
| 4 | 5.198 | 1135.57 | 5.121 | 1359.28 |
| 5 | 5.229 | 645.08 | 5.179 | 690.28 |
| 6 | 5.187 | 1569.12 | 5.131 | 1973.91 |

**Table S3a: The $\beta_x$, $\beta_y$ and $\beta_z$ values (a.u.) of molecules (1-6) as calculated using DFT (CAM-B3LYP functional), employing cc-pVTZ, 6-31G(d,p) and 6-31+G(d,p) basis sets.**

| Molecule | cc-pVTZ | | | 6-31G(d,p) | | | 6-31+G(d,p) | | |
|---|---|---|---|---|---|---|---|---|---|
| | $\beta_x$ | $\beta_y$ | $\beta_z$ | $\beta_x$ | $\beta_y$ | $\beta_z$ | $\beta_x$ | $\beta_y$ | $\beta_z$ |
| 1 | 1130.48 | 2.79 | -0.12 | 1156.26 | -2 | 1.03 | 1415.31 | 0.84 | 0.07 |
| 2 | 482.76 | 16.71 | 4.71 | 427 | 2.59 | -2.23 | 534.84 | 4.65 | 0.24 |
| 3 | -3942.9 | 19.59 | 7.25 | -3992.63 | -1.89 | -68.83 | -4570.73 | 36.30 | 27.64 |
| 4 | 1135.56 | -5.08 | 1.54 | 1234.31 | -0.36 | -4.74 | 1359.28 | 2.25 | 0.21 |
| 5 | -641.9 | 63.5 | 8 | 576.25 | 16.56 | 0.62 | -680.61 | 115.14 | -0.2 |
| 6 | 1568.47 | -28.22 | 35.57 | -1714.75 | -3.16 | 11.95 | -1973.91 | 0.12 | 1.27 |

**Table S3b: The $\beta_x$, $\beta_y$ and $\beta_z$ values (a.u.) of molecules (2-4) as calculated using MP2 theory, employing 6-31G(d,p) and 6-31+G(d,p) basis sets.**

| Molecule | 6-31G(d,p) | | | 6-31+G(d,p) | | |
|---|---|---|---|---|---|---|
| | $\mu_x$ | $\mu_y$ | $\mu_z$ | $\mu_x$ | $\mu_y$ | $\mu_z$ |
| 2 | 55.39 | -1.36 | -31.90 | 140.52 | 47.26 | -48.04 |
| 3 | -4001.11 | 52.15 | -84.14 | -4331.45 | 176.95 | -0.27 |
| 4 | 849.26 | 4.09 | -17.57 | 919.44 | -1.07 | -33.86 |

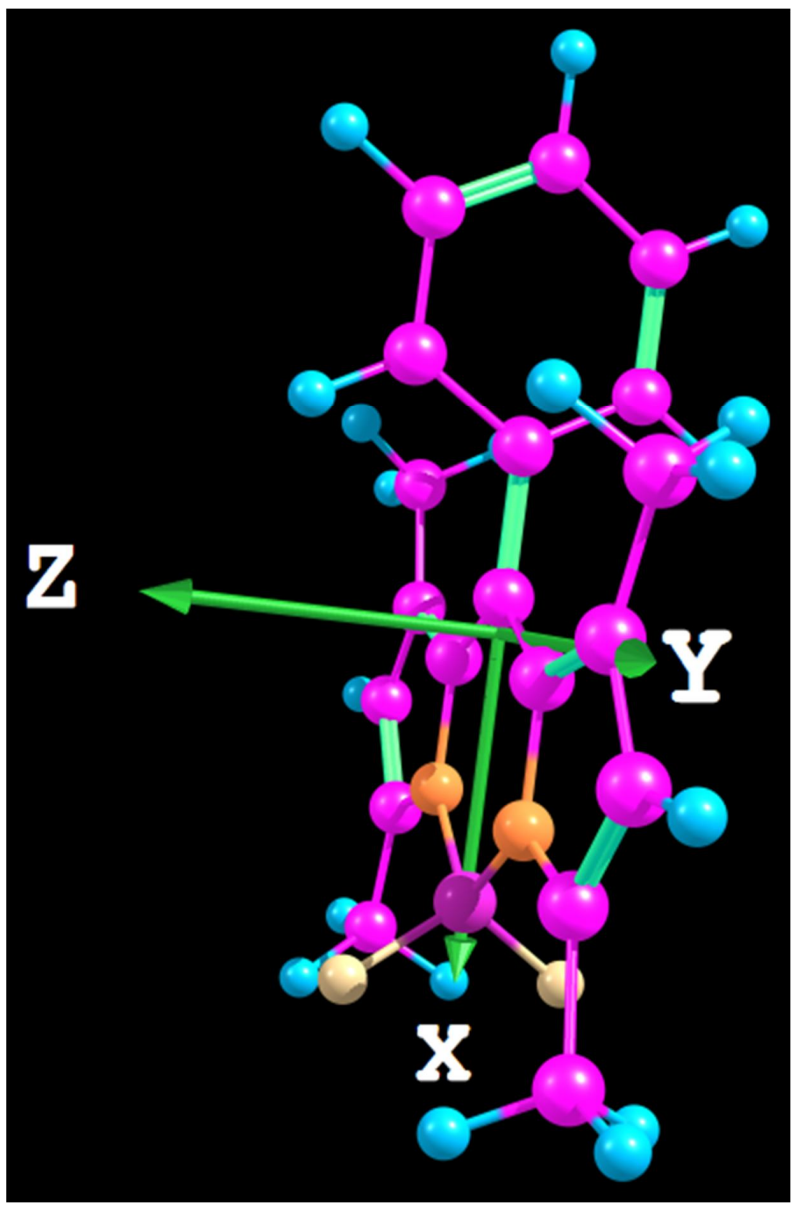

**Figure S1(1)**

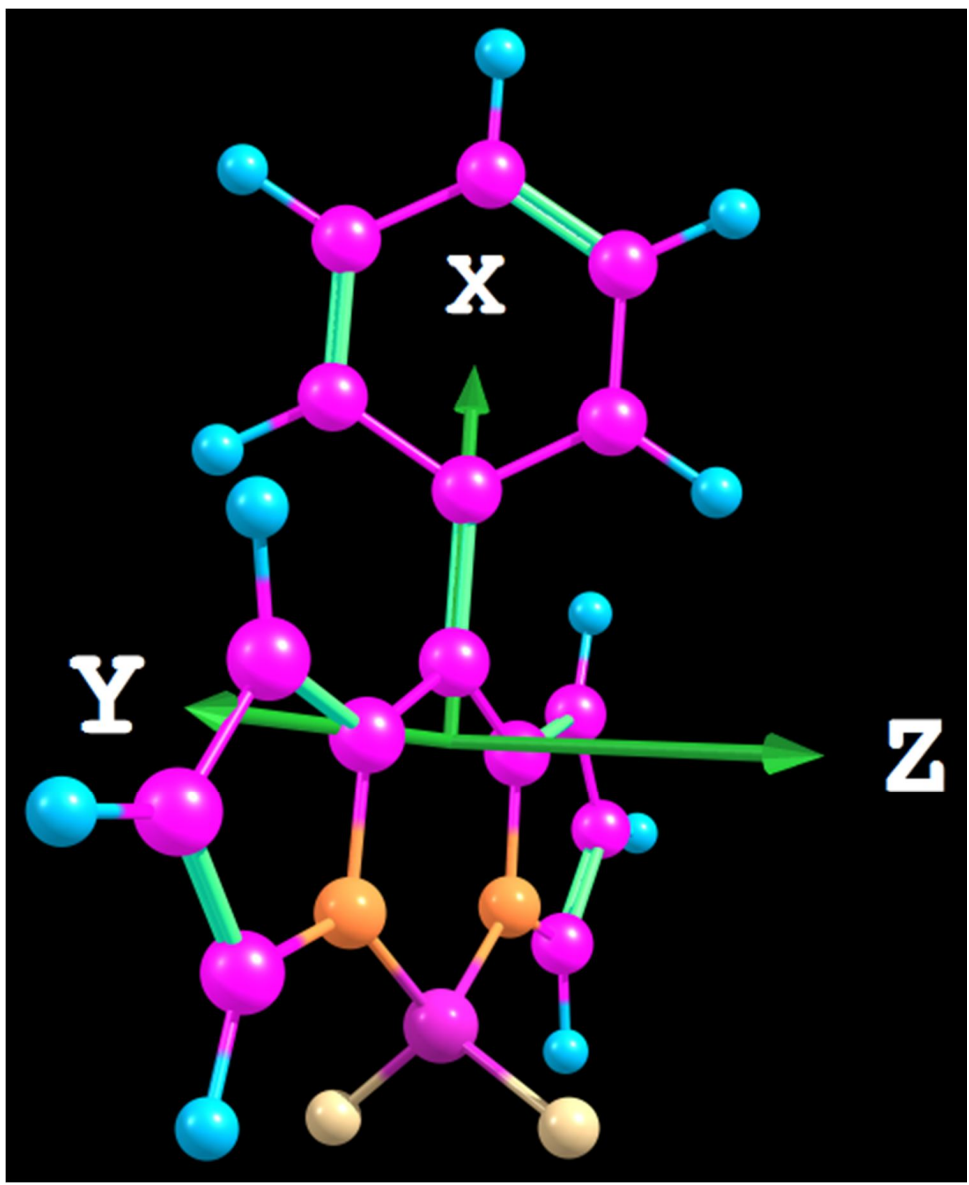

**Figure S1(2)**

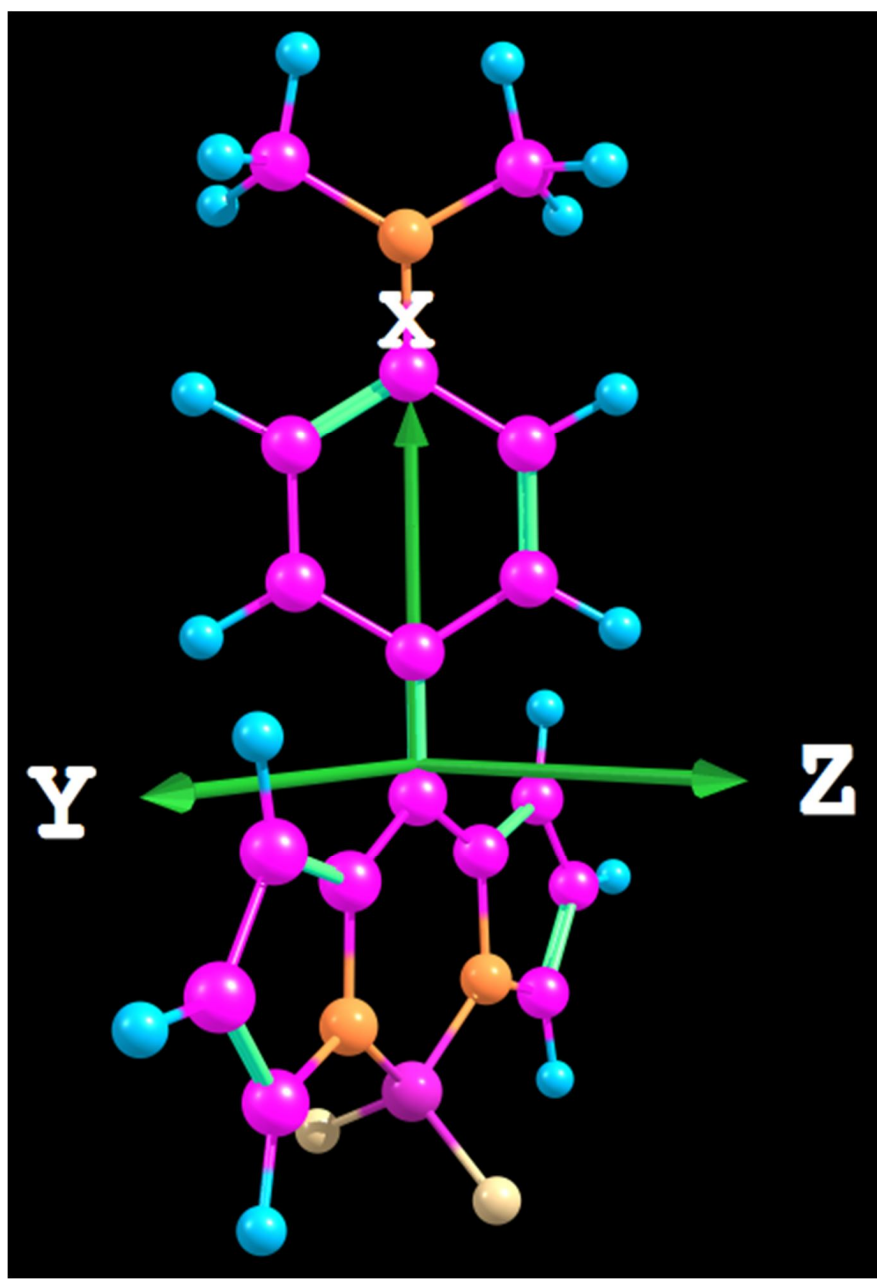

**Figure S1(3)**

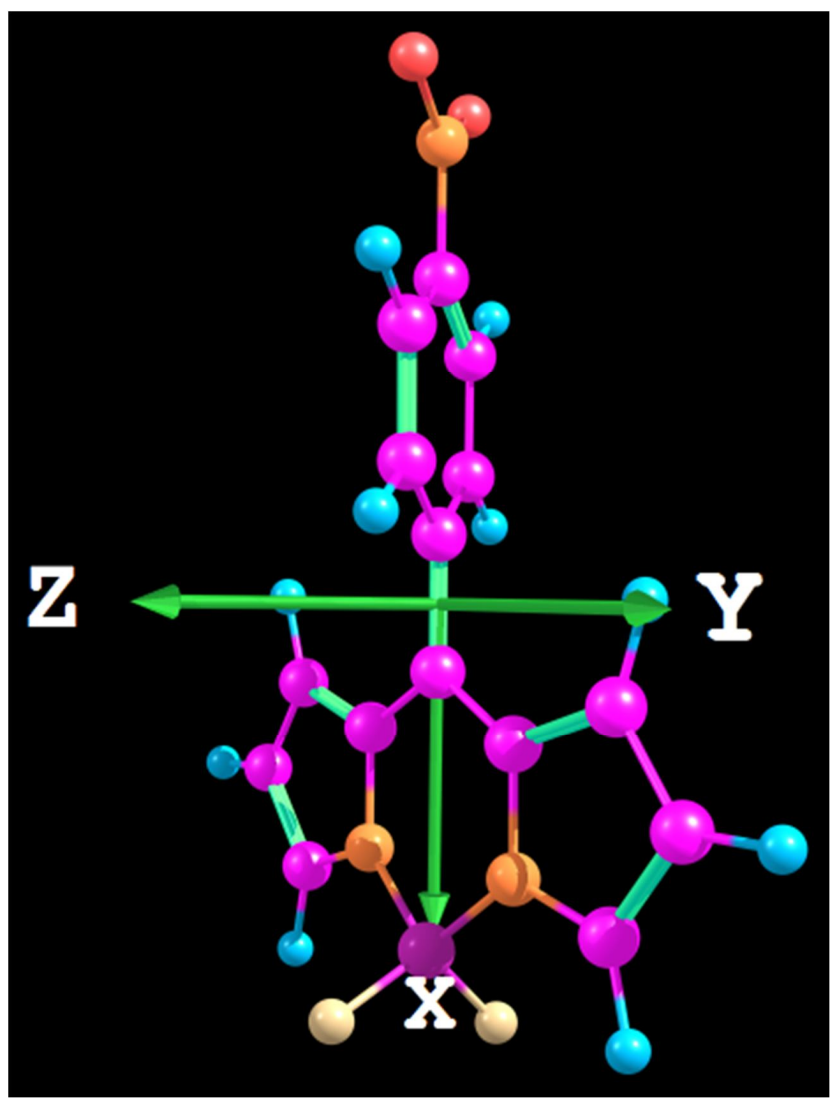

**Figure S1(4)**

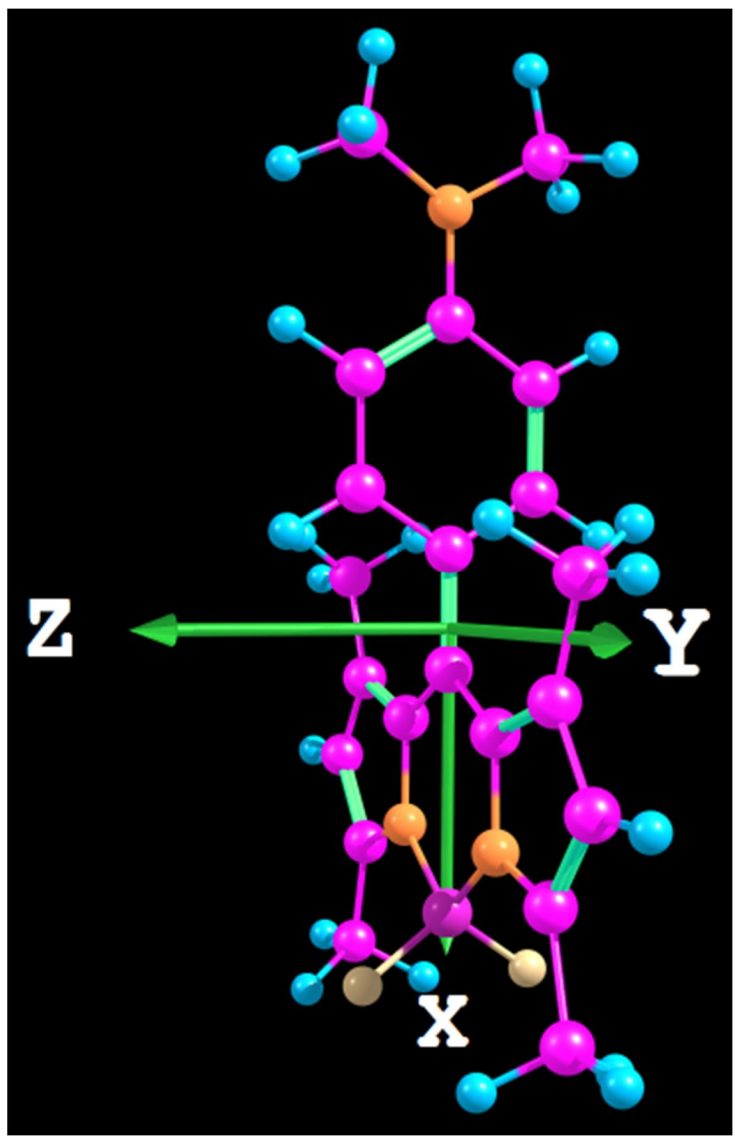

**Figure S1(5)**

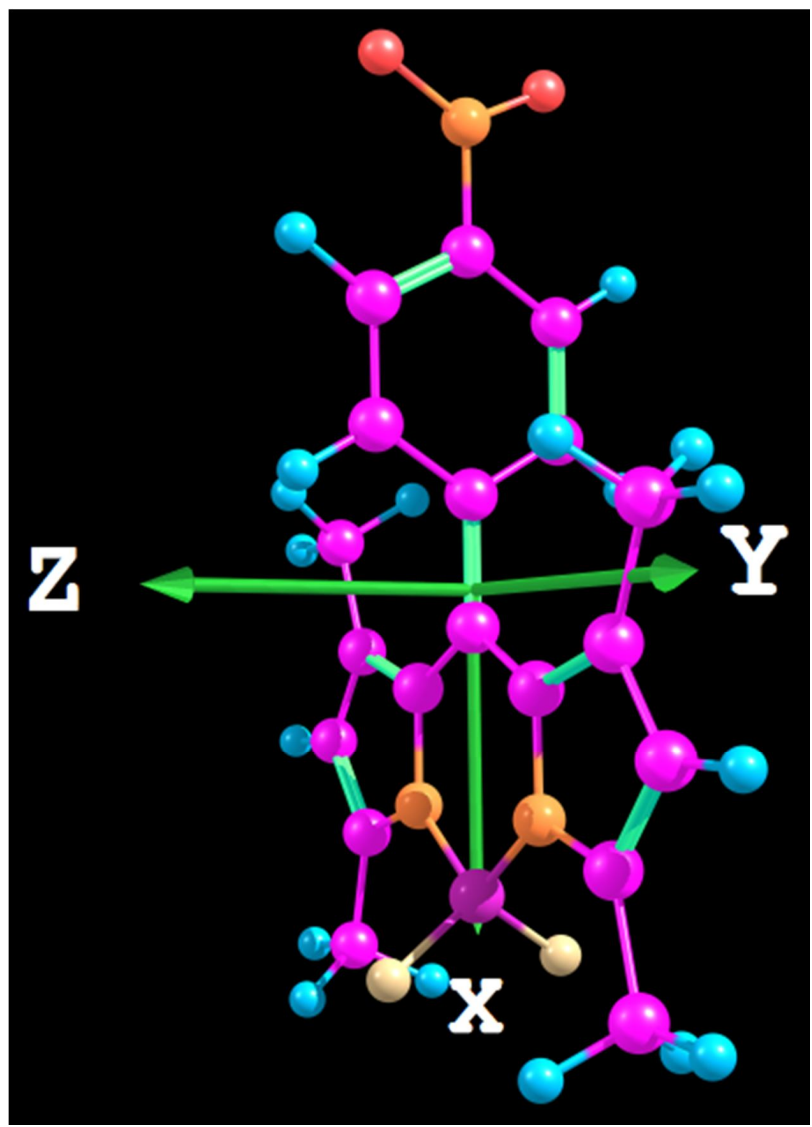

**Figure S1(6)**

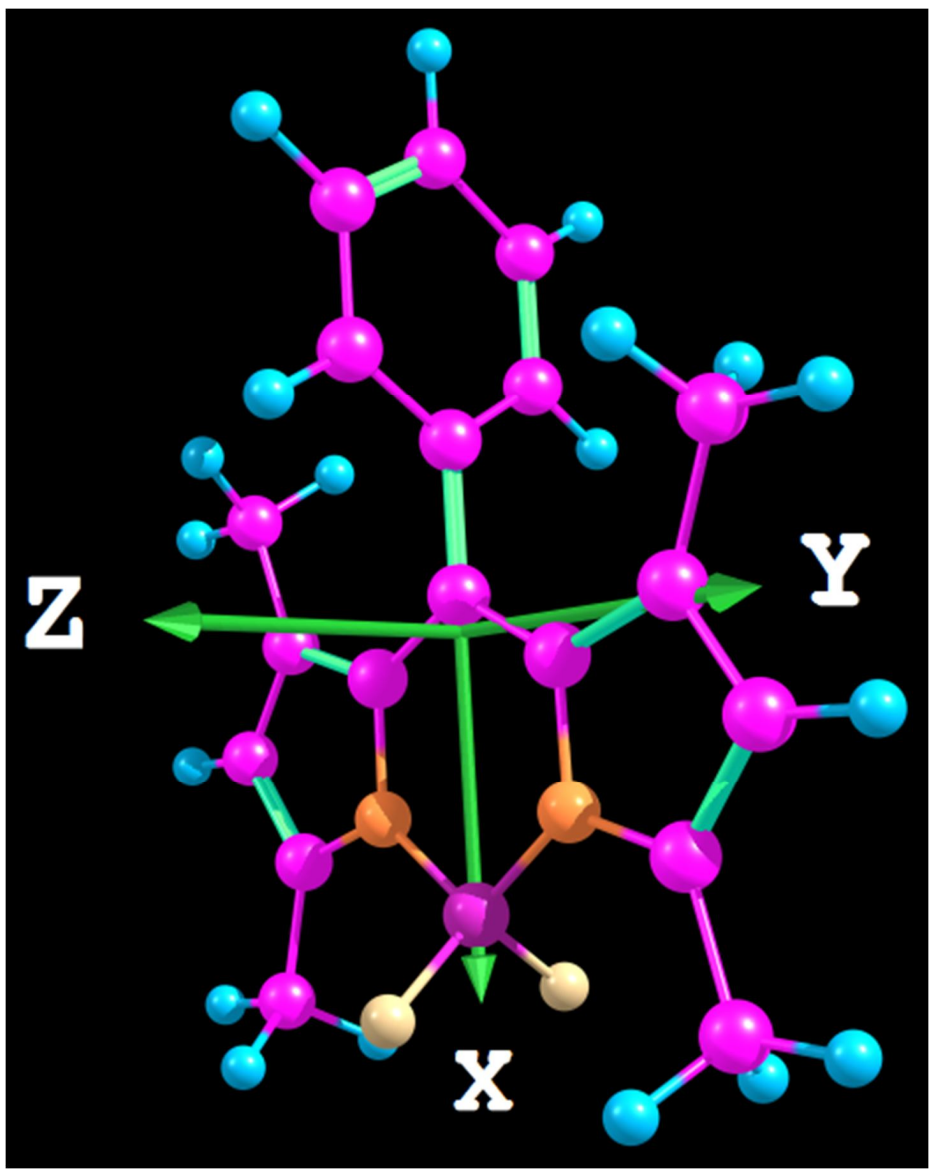

**Figure S2(1)**

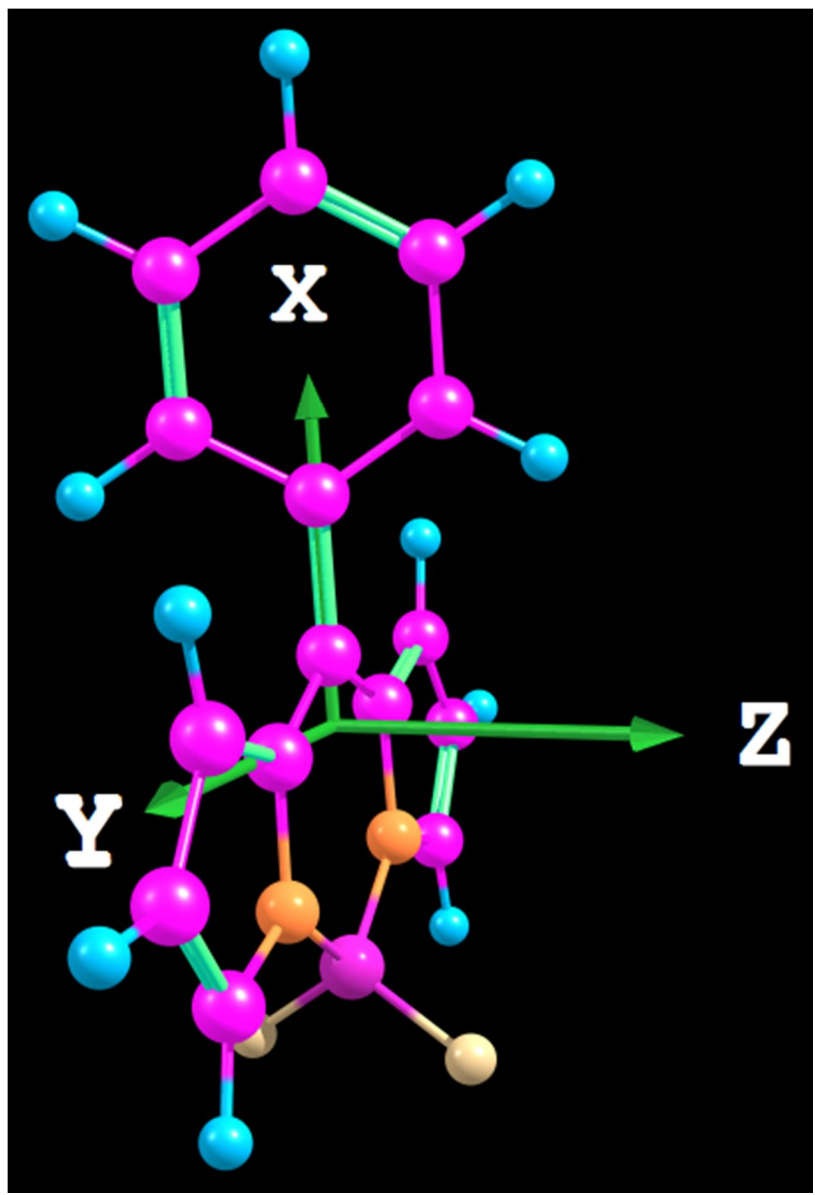

**Figure S2(2)**

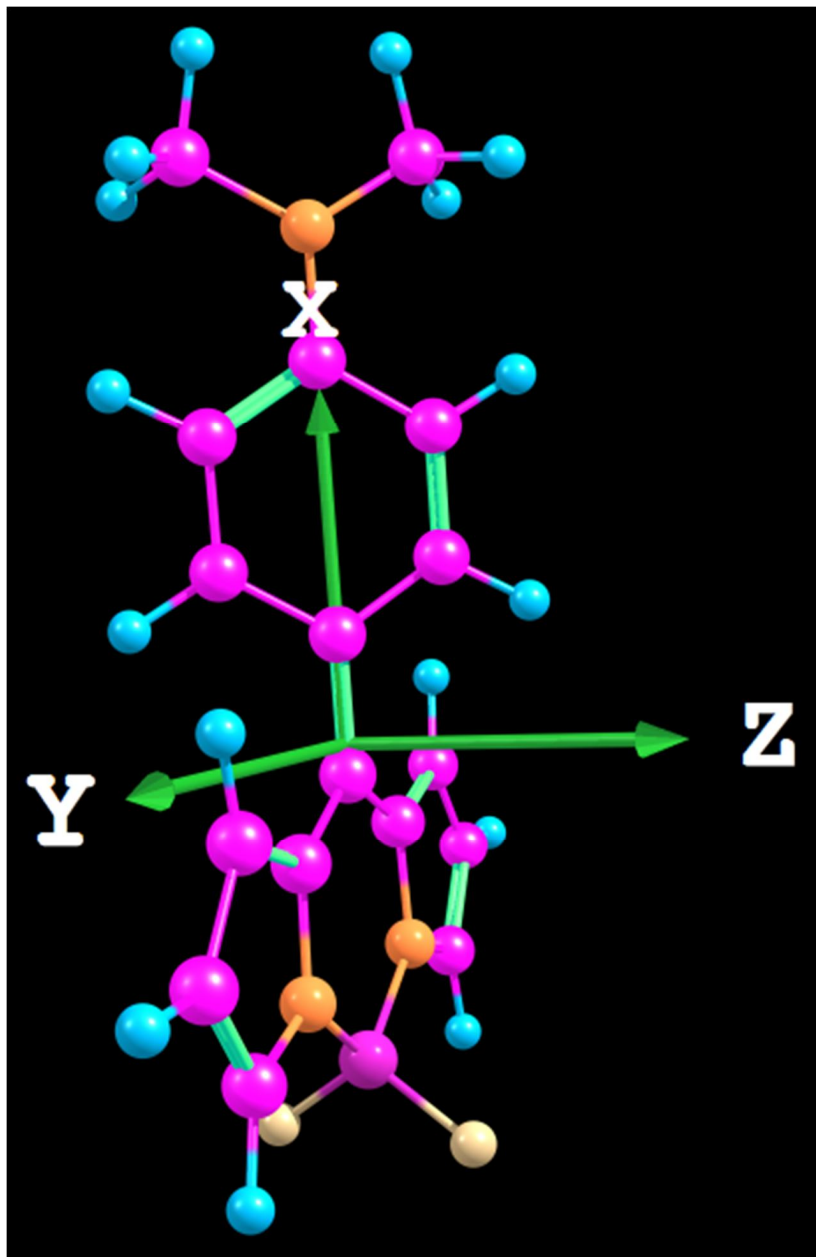

**Figure S2(3)**

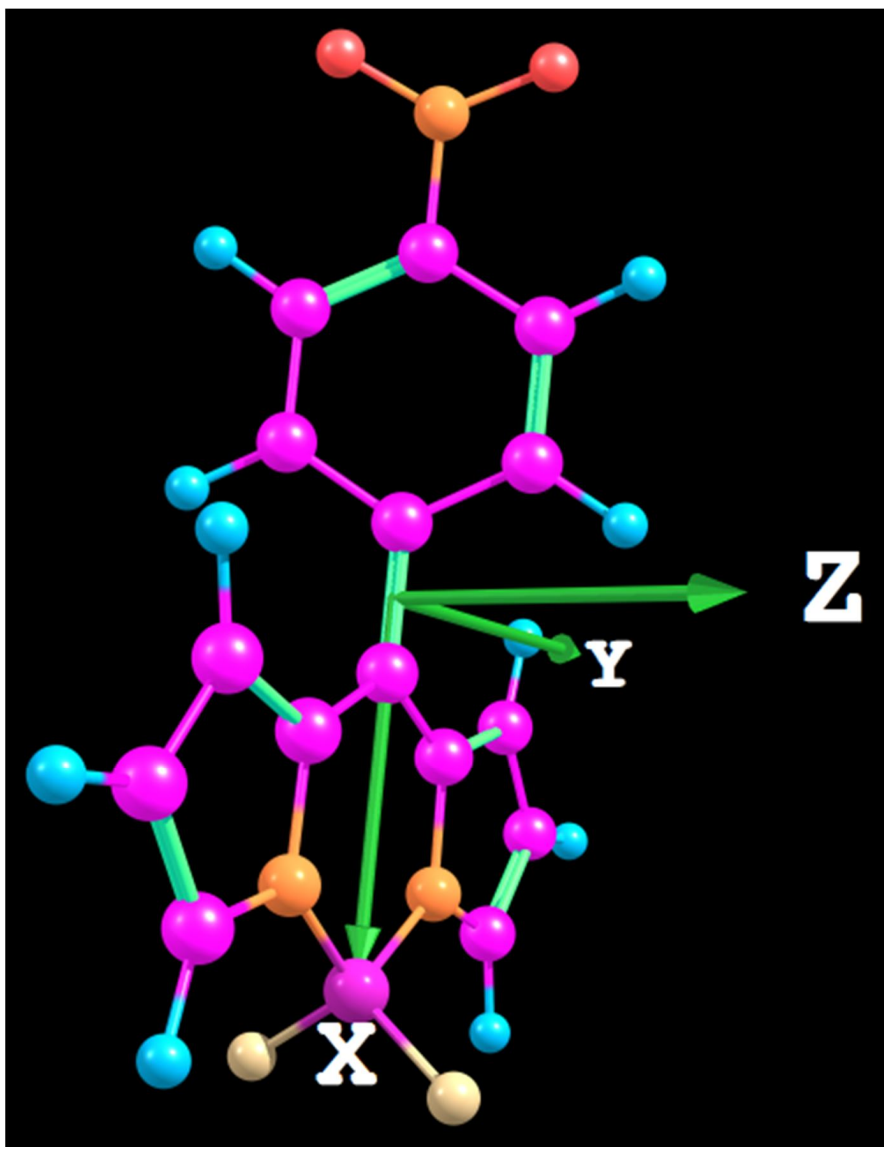

**Figure S2(4)**

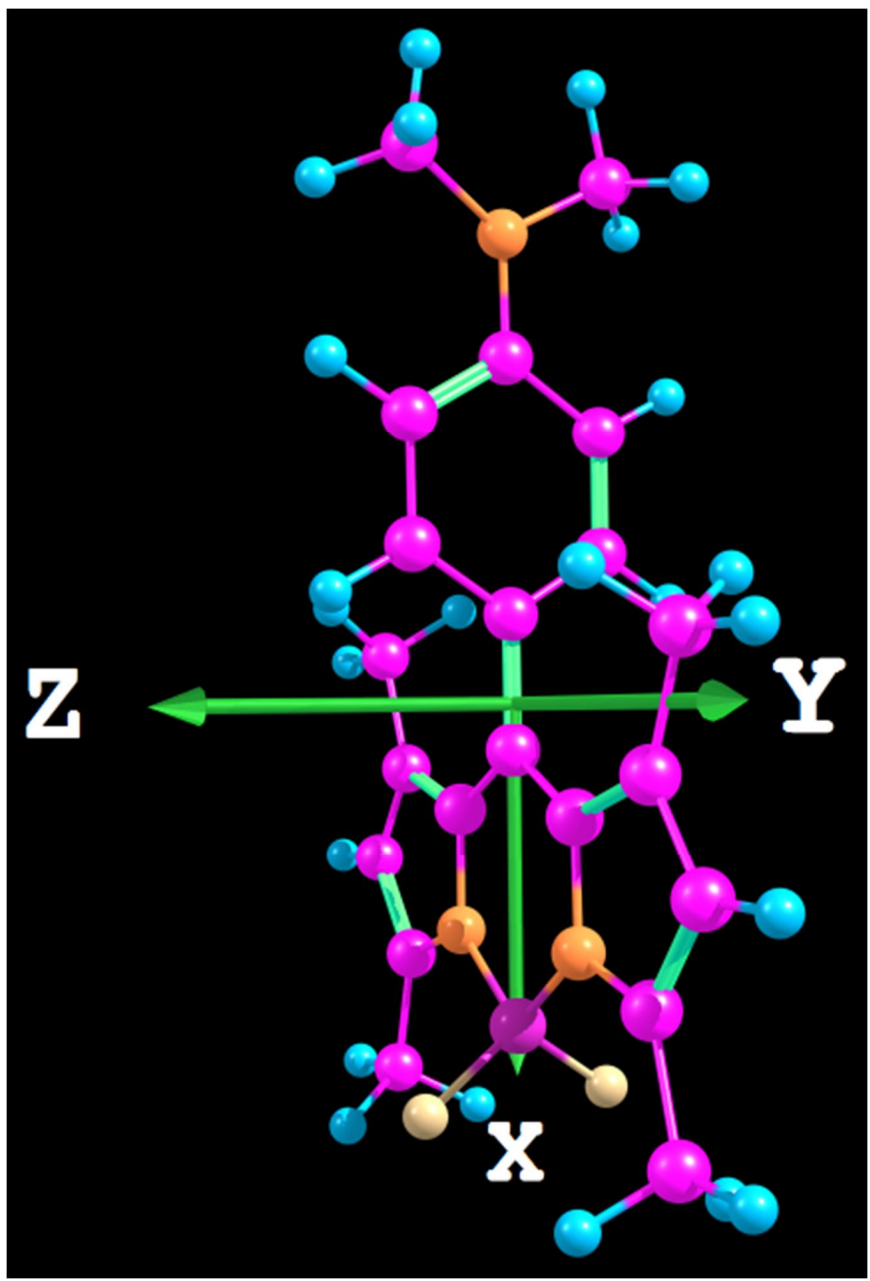

**Figure S2(5)**

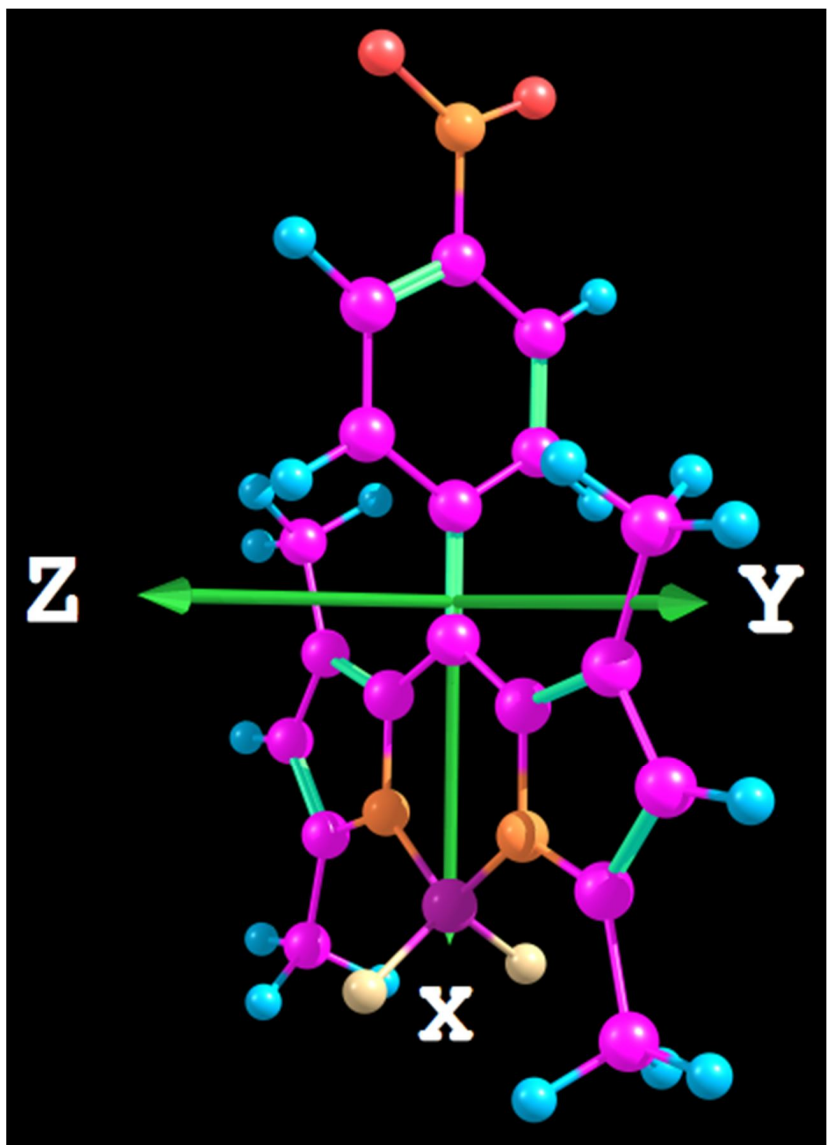

**Figure S2(6)**

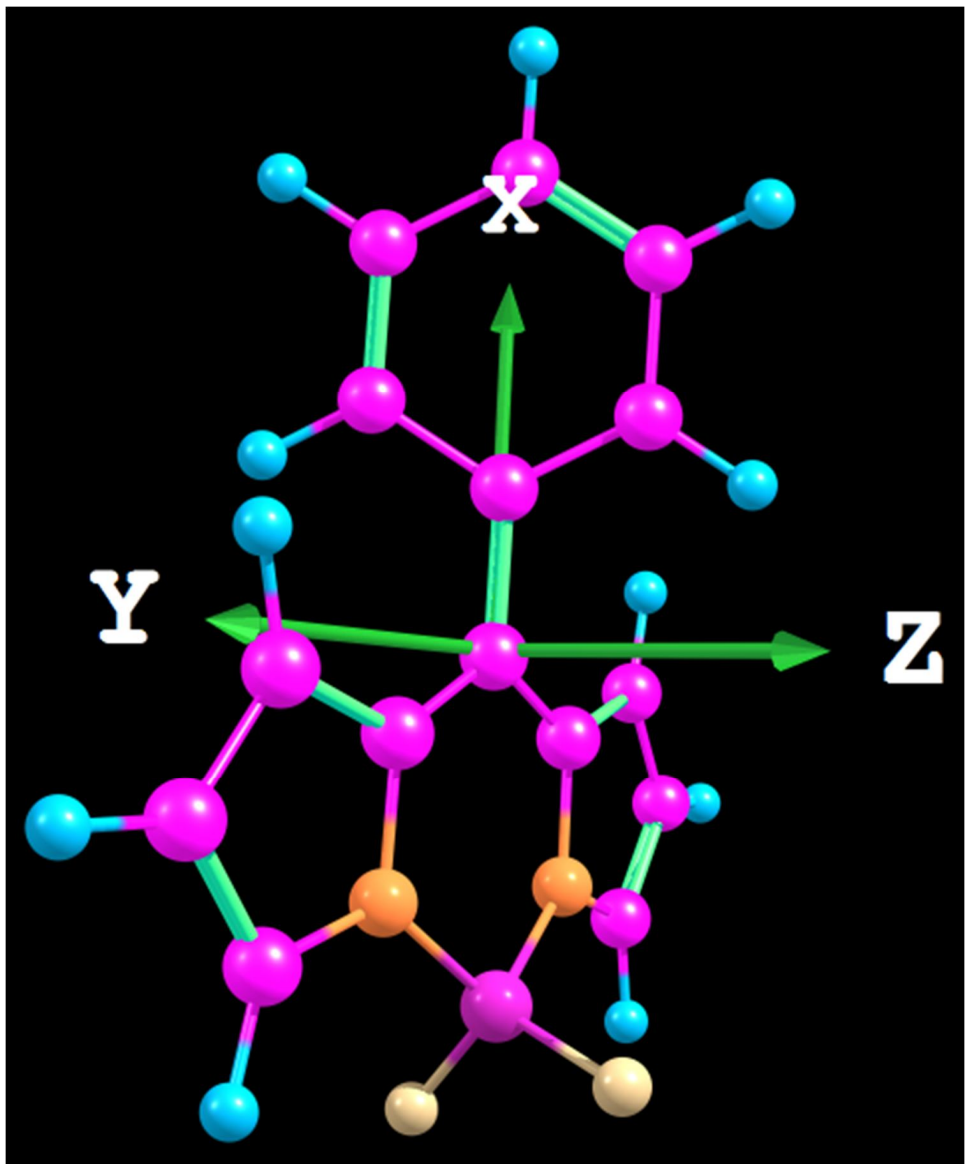

**Figure S3 (2)**

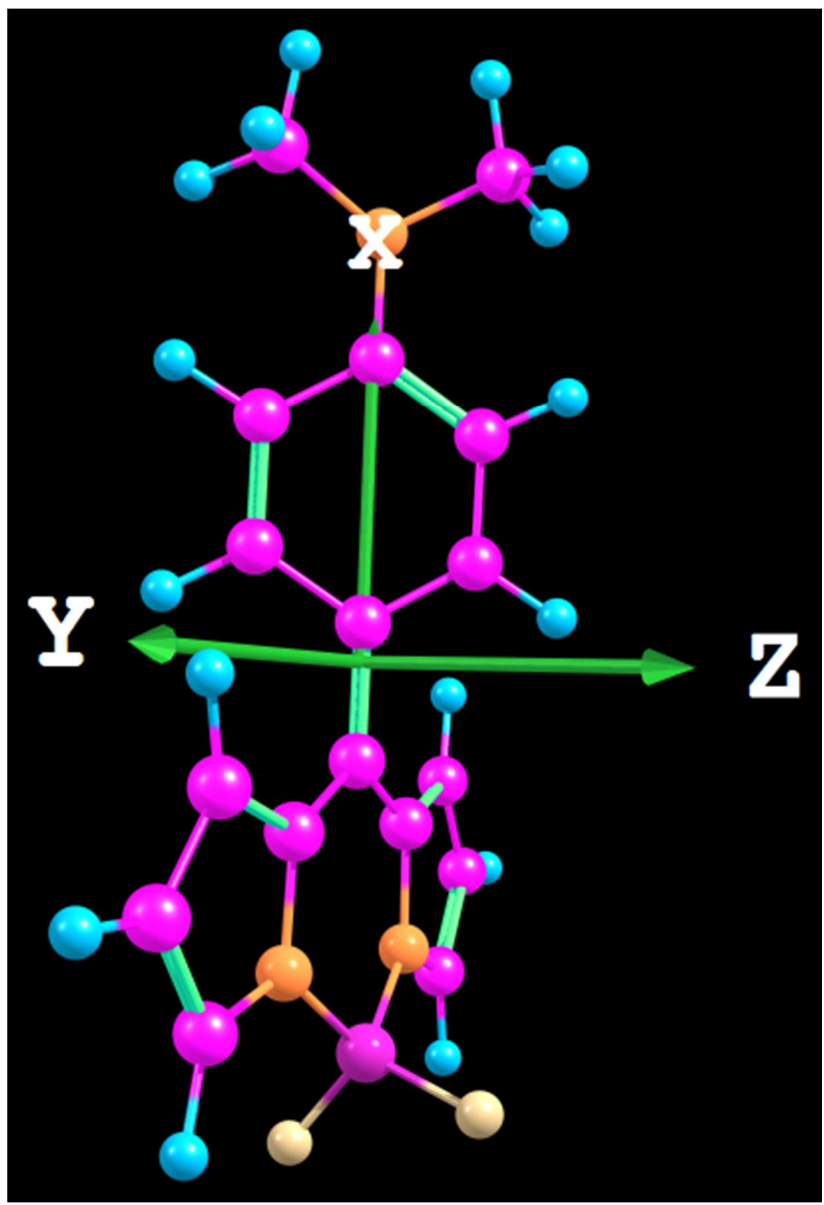

**Figure S3 (3)**

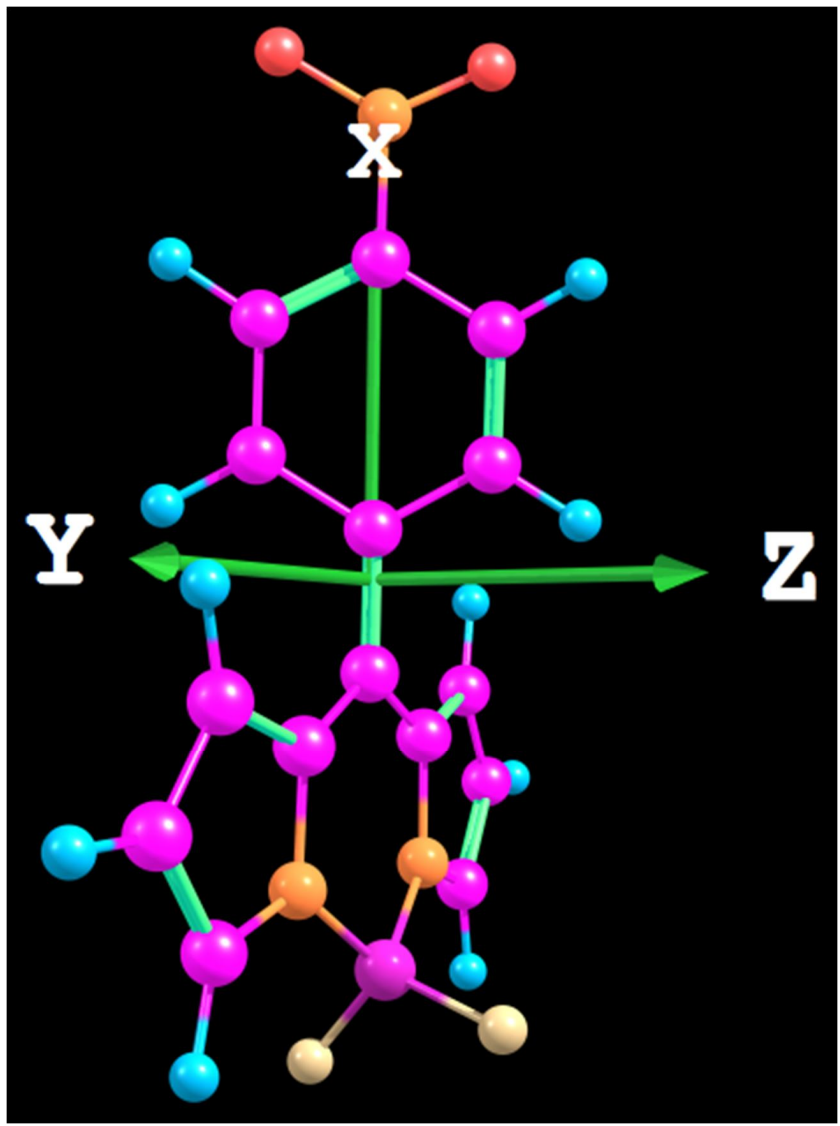

**Figure S3 (4)**

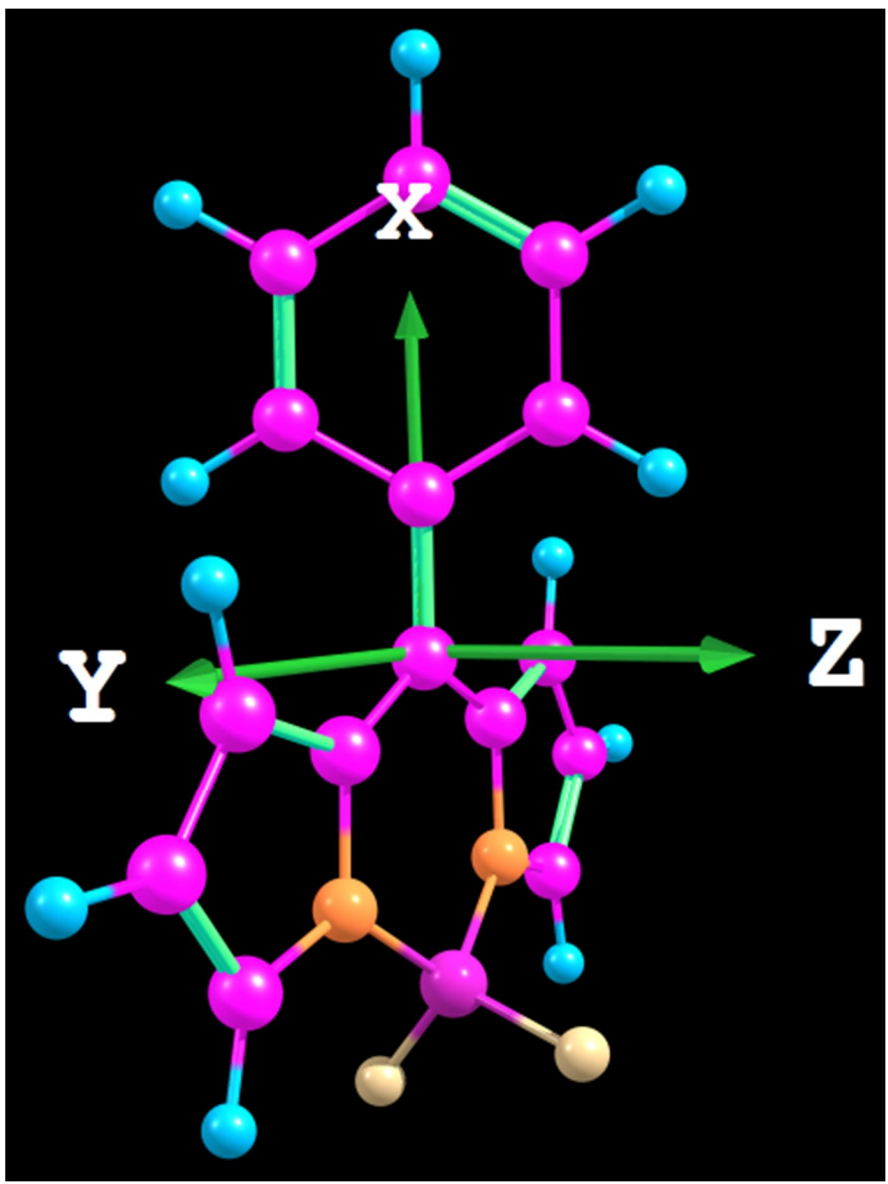

**Figure S4 (2)**

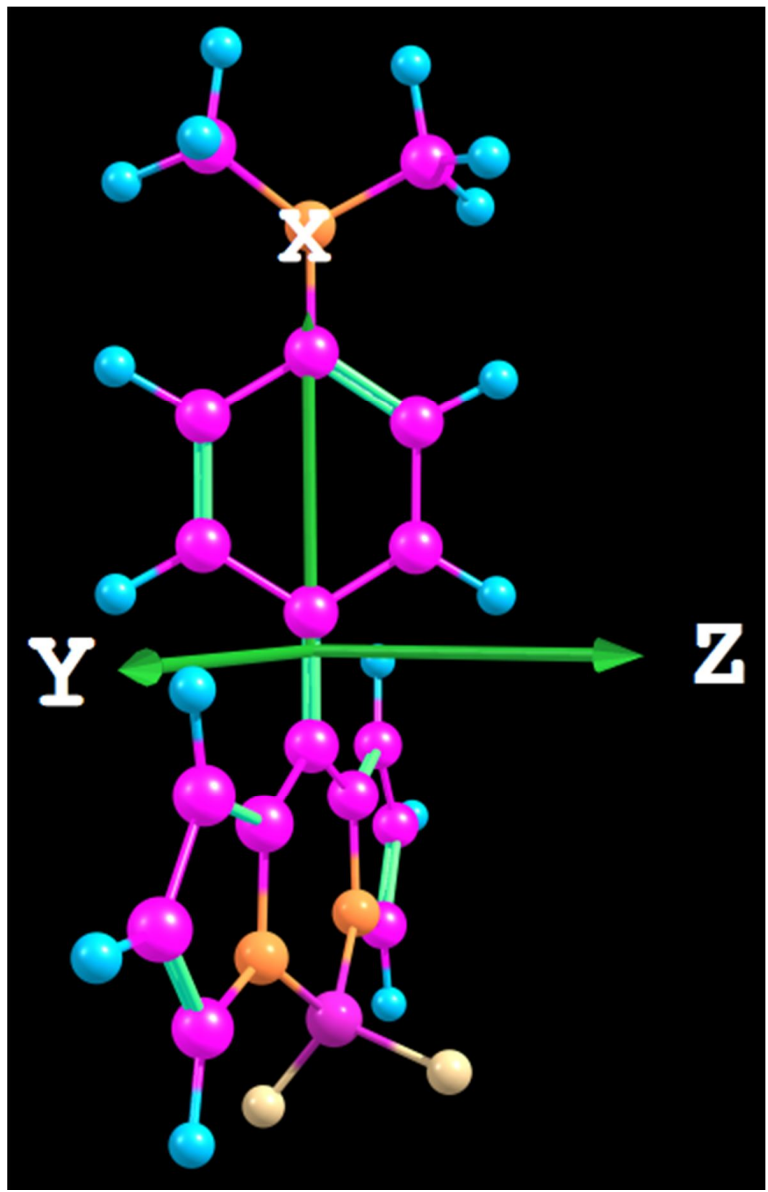

**Figure S4 (3)**

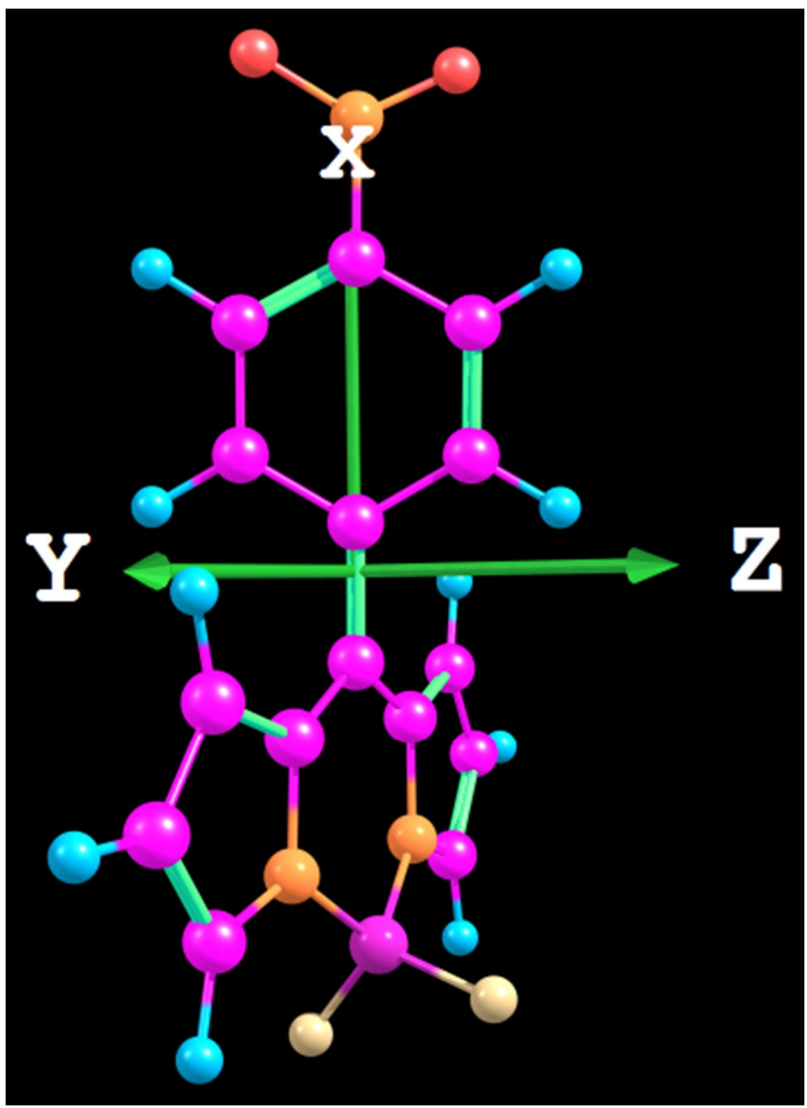

**Figure S4 (4)**